\documentclass[superscriptaddress,altaffilletter,amsmath,amssymb,aps,twocolumn,floatfix]{revtex4-1}
\usepackage[english]{babel}
\usepackage[colorinlistoftodos]{todonotes}
\usepackage{multirow}
\usepackage{comment}
\usepackage{relsize}
\usepackage{float}
\usepackage{textcomp}
\usepackage{hyperref}
\usepackage[justification=justified]{subcaption}
\usepackage[justification=justified, figurename=FIG.]{caption}
\usepackage[compact]{titlesec}
\titlespacing{\section}{0pt}{3ex}{2ex}
\titlespacing{\subsection}{0pt}{2ex}{2ex}
\titlespacing{\subsubsection}{0pt}{1.5ex}{1.5ex}

\DeclareCaptionLabelFormat{UC}{\MakeUppercase{#1} #2}
\usepackage{xcolor}
\usepackage{graphicx}% Include figure files
\usepackage{epstopdf}
\usepackage{color}
\usepackage{dcolumn}% Align table columns on decimal point
\usepackage{bm}% bold math
\usepackage{hyperref}% add hypertext capabilities
\usepackage[mathlines]{lineno}% Enable numbering of text and display math
\usepackage{threeparttable} % for more complex tables
\usepackage{booktabs} % for the midrule command et al
\usepackage{siunitx}

\begin{document}

\title{Measurement of the Gamma Ray Background in the Davis Cavern at the Sanford Underground Research Facility}  
%1
% 1 
\author{D.S.~Akerib}
\affiliation{SLAC National Accelerator Laboratory, Menlo Park, CA 94025-7015, USA}
\affiliation{Kavli Institute for Particle Astrophysics and Cosmology, Stanford University, Stanford, CA  94305-4085 USA}

% 2 
\author{C.W.~Akerlof}
\affiliation{University of Michigan, Randall Laboratory of Physics, Ann Arbor, MI 48109-1040, USA}

% 3 
\author{S.K.~Alsum}
\affiliation{University of Wisconsin-Madison, Department of Physics, Madison, WI 53706-1390, USA}

% 4 
\author{N.~Angelides}
\affiliation{University College London (UCL), Department of Physics and Astronomy, London WC1E 6BT, UK}

% 5 
\author{H.M.~Ara\'{u}jo}
\affiliation{Imperial College London, Physics Department, Blackett Laboratory, London SW7 2AZ, UK}

% 6 
\author{J.E.~Armstrong}
\affiliation{University of Maryland, Department of Physics, College Park, MD 20742-4111, USA}

% 7 
\author{M.~Arthurs}
\affiliation{University of Michigan, Randall Laboratory of Physics, Ann Arbor, MI 48109-1040, USA}

% 8 
\author{X.~Bai}
\affiliation{South Dakota School of Mines and Technology, Rapid City, SD 57701-3901, USA}

% 9 
\author{J.~Balajthy}
\affiliation{University of California, Davis, Department of Physics, Davis, CA 95616-5270, USA}

% 10 
\author{S.~Balashov}
\affiliation{STFC Rutherford Appleton Laboratory (RAL), Didcot, OX11 0QX, UK}

% 11 
% ...

\collaboration{The LUX-ZEPLIN (LZ) Collaboration}

\date{\today}
\begin{abstract}
\noindent 
%\linenumbers % Commence numbering lines
Deep underground environments are ideal for low background searches due to the attenuation of cosmic rays by passage through the earth. However, they are affected by backgrounds from $\gamma$-rays emitted by $^{40}$K and the $^{238}$U and $^{232}$Th decay chains in the surrounding rock. The LUX-ZEPLIN (LZ) experiment will search for dark matter particle interactions with a liquid xenon TPC located within the Davis campus at the Sanford Underground Research Facility, Lead, South Dakota, at the 4,850-foot level. In order to characterise the cavern background, in-situ $\gamma$-ray measurements were taken with a sodium iodide detector in various locations and with lead shielding. The integral count rates (0--3300~keV) varied from 596~Hz to 1355~Hz for unshielded measurements, corresponding to a total flux from the cavern walls of $1.9\pm0.4$~$\gamma~$cm$^{-2}$s$^{-1}$. The resulting activity in the walls of the cavern can be characterised as $220\pm60$~Bq/kg of $^{40}$K, $29\pm15$~Bq/kg of $^{238}$U, and $13\pm3$~Bq/kg of $^{232}$Th.

\end{abstract}

\maketitle

\section{\label{sec:Introduction}Introduction}
\noindent 
Direct searches for Weakly Interacting Massive Particles (WIMPs) as candidates for dark matter involve looking for an interaction of a WIMP with an atomic nucleus, a process not yet observed. Recent experiments, utilising xenon as the target in dual-phase noble liquid time projection chambers (TPCs), have led the way in setting ever more stringent limits on WIMP properties~\cite{Akerib:2016vxi, Cui:2017,Aprile:2018}. In order to probe smaller and smaller WIMP-nucleon interaction cross sections, such rare-event searches demand extremely low background event rates in the  region of interest. The LUX-ZEPLIN (LZ) experiment, a dark matter experiment presently under construction~\cite{mount:2017} at the Sanford Underground Research Facility (SURF), is projected to reach unprecedented sensitivity, excluding spin-independent scattering cross sections above $1.6\times10^{-48}$~cm$^2$ for a 40~GeV/c$^2$ WIMP~\cite{LZ:2018}.

LZ will feature a dual-phase xenon TPC containing 7~tonnes of active xenon inside a radiopure titanium cryostat~\cite{akerib:2017}. Signals from electron and nuclear recoils induced by $\gamma$-rays, electrons, neutrinos, neutrons, and potentially WIMPs are collected by a total of 494 photomultiplier tubes (PMTs). A two-component veto system rejects any particles scattering in both the TPC and veto detectors and characterises the general radioactive backgrounds for LZ.  The veto system consists of an instrumented liquid xenon skin between the TPC and inner cryostat, primarily to detect scattered $\gamma$-rays, and a near-hermetic 17.3~T Gd-loaded liquid scintillator detector system known as the Outer Detector (OD) surrounding the outer cryostat. The main purpose of the OD is to veto neutrons; Gd has an extremely high thermal neutron capture cross section, making neutrons easily detectable via the post-capture $\gamma$-ray cascade. A high neutron detection efficiency in the OD is essential as an approximately 1~MeV neutron characteristic of those born via natural radioactivity can produce a nuclear recoil in the same energy window as a WIMP. Maintaining a low background rate inside the OD from sources external to the detector, such as the $\gamma$-rays from natural radioactivity in the cavern, is essential for its role as a veto. This is due to considerations of the false veto rate and the amount of excluded data, both of which will increase with the rate in the OD. 

SURF is located at the former Homestake Gold Mine in Lead, South Dakota. The Davis cavern, the future home of the LZ experiment, is located 4,850~feet (4,300 m.w.e.~\cite{lesko:2015sma}) underground, and was home to LZ's predecessor LUX, a 250~kg xenon detector which set world leading constraints on WIMP-nucleon scattering cross-sections~\cite{Akerib:2016vxi,Akerib:2016lao}. At this depth, the cosmic ray flux is reduced by a factor of $10^{6}$ compared with that at sea level~\cite{cherry:1983dp,gray:2010nc,majorana:2017}; however, when underground a background from intrinsic radioactivity in the cavern rock must be considered. LZ will be housed within a water tank of height 591~cm and radius 381~cm which provides additional shielding from this radioactivity. This water tank previously housed the LUX detector and was instrumented with PMTs to act as a water Cherenkov muon veto~\cite{akerib:2012ys}. Further shielding is provided by 6 octagonal plates of 5~cm thickness inlaid in the floor beneath the water tank.  This shield has the shape of an inverted pyramid beneath the centre of the water tank, directly below where the xenon target is placed for both LUX and LZ. The tank and the pyramid sit atop a layer of gravel that extends as deep as the pyramid and to the radius of the water tank. 
\begin{figure}[h]
\begin{minipage}{0.75\linewidth}
\includegraphics[width=1\linewidth, trim =0 0 0 0]{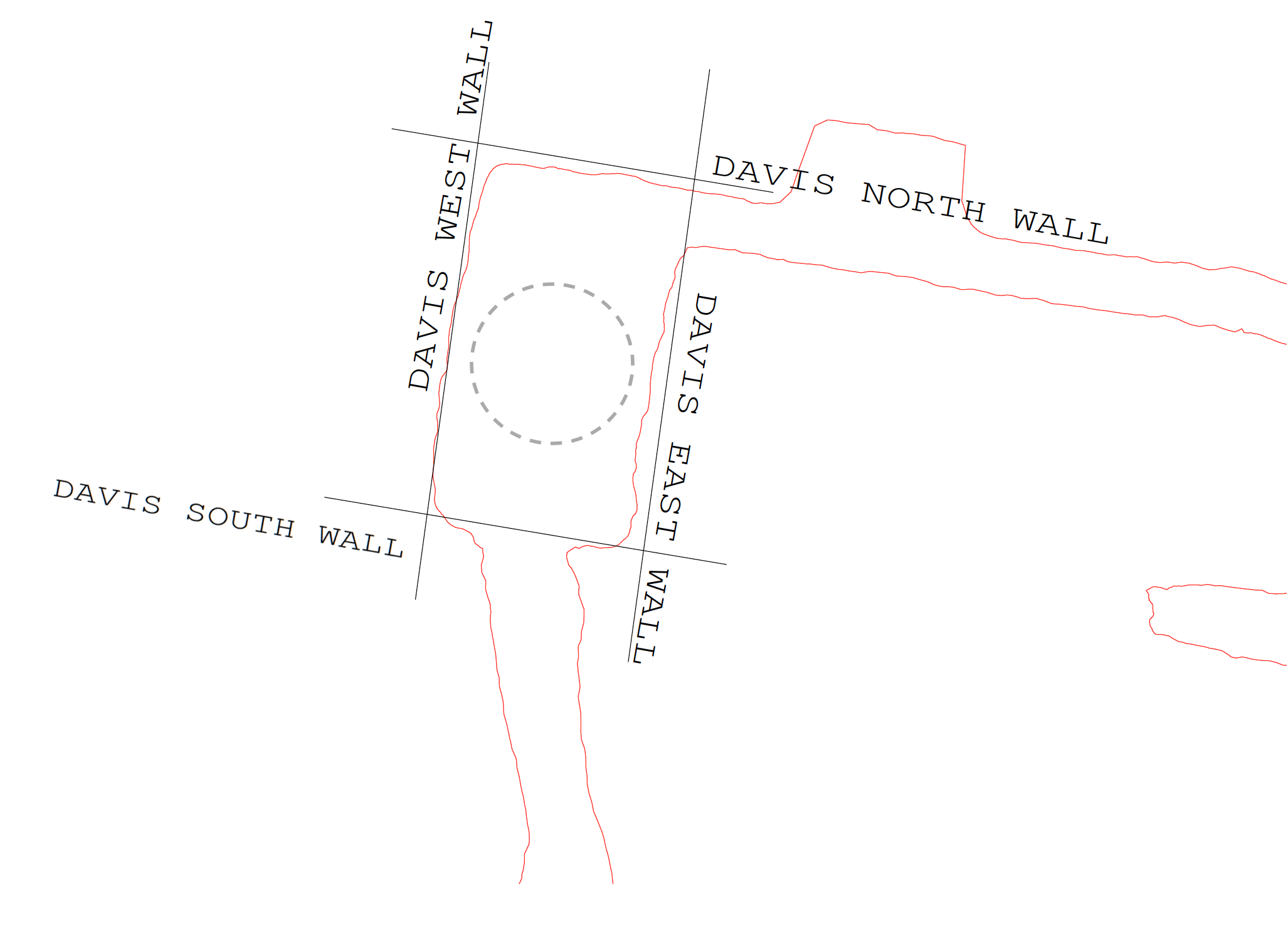}
\subcaption{\label{subfig:a}}
\end{minipage} \\\vspace{10pt}
\begin{minipage}{0.32\linewidth}
\includegraphics[height=2cm, trim =200 200 200 200]{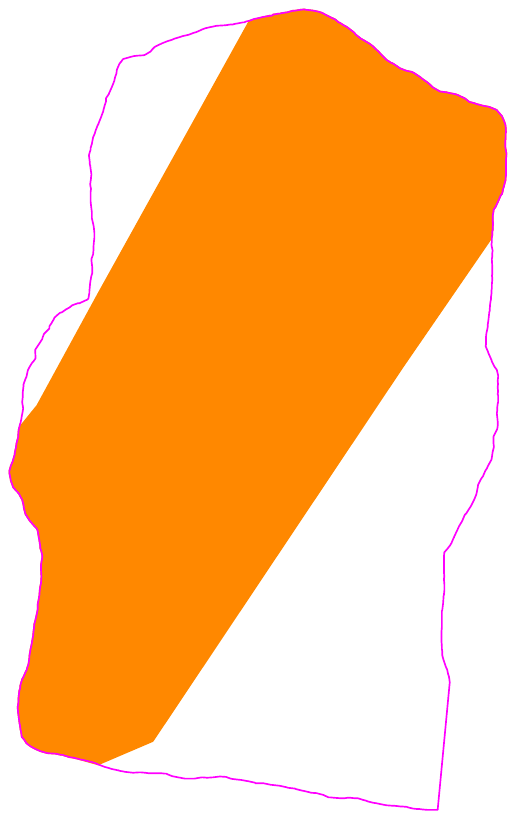} 
\subcaption{\footnotesize Floor - 64\% rhyolite \label{subfig:b}}
\end{minipage} 
\begin{minipage}{0.32\linewidth}
  \includegraphics[height=2cm, trim =250 260 200 200]{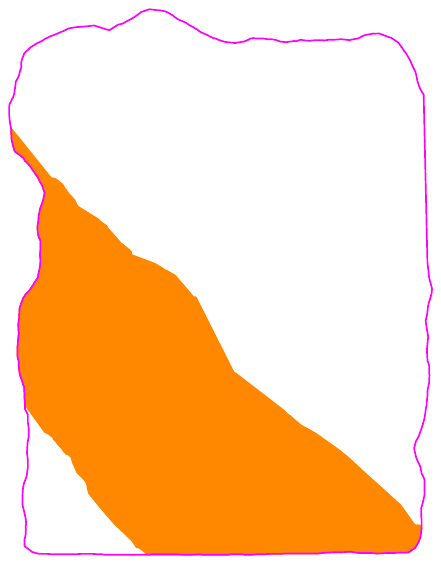}
\subcaption{\footnotesize North wall - 36\% rhyolite \label{subfig:c}}
\end{minipage}
\begin{minipage}{0.32\linewidth}
\includegraphics[height=2cm, trim =200 200 200 200]{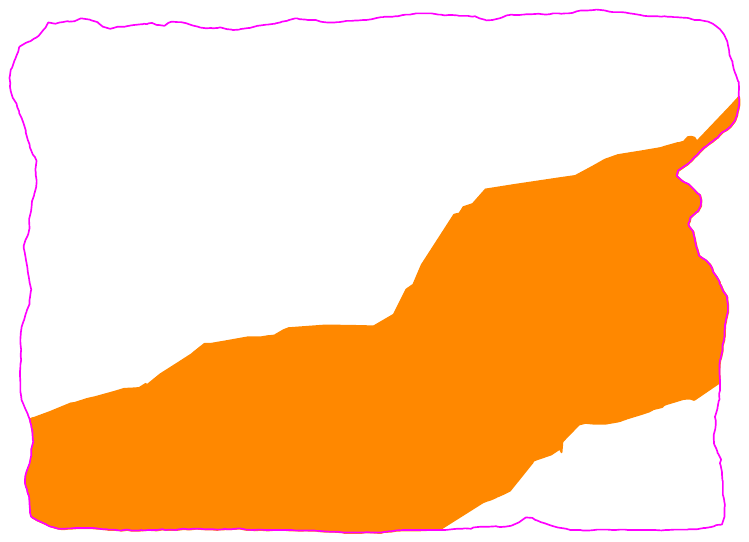}
\subcaption{\footnotesize West wall - 42\% rhyolite \label{subfig:d}}
\end{minipage}
\caption{\small Schematic of the Davis cavern is shown in \ref{subfig:a}, with the naming convention of the walls and the location of the water tank marked in dashed grey. Expected location of the rhyolite intrusion is shown in orange in \ref{subfig:b}--\ref{subfig:d} with the relative contribution of rhyolite given in the caption. The other walls are not shown; the ceiling is estimated at 0\% rhyolite, the south wall at 5\% and the east at 2\%~\cite{hart:2010}.  \label{fig:rhyolite}}
\end{figure}
Geological and radiometric surveys of the Homestake mine indicate that most rock at the 4,850 level is of the Homestake formation, a metamorphic rock of relatively low uranium and thorium content~\cite{heise:2015}. Additionally, rhyolite intrusions in the rock have been identified; rhyolite is  an igneous, volcanic, silica-rich rock, with higher natural radioactivity. The relative amount of rhyolite intrusions within the Davis cavern has been estimated; the intrusion is known to pass below the cavern and diagonally across both the north wall and west wall (see Figs.~\ref{subfig:b}--\ref{subfig:d}). 
A layer of sprayed concrete (shotcrete) of average thickness 12.7~cm lines the walls and ceiling of the cavern, but the thickness is known to vary by a factor of two. The floor consists of 15~cm of low-radioactivity concrete; the exception to this is within two rooms at the north end of the lower level of the cavern, originally used for low background HPGe measurements, known as the counting rooms, where the floor is 30~cm thick.
The radiological contents (uranium, thorium, potassium) of rock formations and construction materials - measured with high purity germanium (HPGe) screening at the time of construction - are shown in Table~\ref{tab:radioactivities}; some variation of radioactivity between all material samples is observed and so both the average values and range are given. Table~\ref{tab:radioactivities} also contains two recent measurements of samples collected at the time of the measurement described in this paper, shown with uncertainties in the bottom two rows.

The $^{238}$U and $^{232}$Th chains contain a series of $\alpha$ and $\beta$-decays which often lead to $\gamma$-ray emission from excited states of the daughters. For the entire uranium and thorium decay chains, on average 2.2 and 2.7 $\gamma$-rays are expected respectively~\cite{nudat:1996}, if secular equilibrium is assumed. Additionally, $^{40}$K emits a 1461~keV $\gamma$-ray with a branching ratio of about 10\%. Because of the possible high levels of these isotopes in both rock formations and construction materials, characterisation of the $\gamma$-ray background in underground facilities is a standard measurement and has been performed at Gran Sasso National Laboratory~\cite{gransasso:2012}, the Modane Underground Laboratory~\cite{modane:2012}, the Boulby Underground Laboratory~\cite{boulby:2013} and the China Jinping Underground Laboratory~\cite{jinping:2014}.

\begin{table}
\centering
\caption{\small Measured activities from radioassay of rock, shotcrete and gravel samples. The first four materials were radioassayed during construction of the Davis cavern and give the average and range for several samples~\cite{smith:2007}. When not stated, overall uncertainties are estimated to be 10--20\%. The final two samples, shotcrete taken from near the entrance to the cavern and gravel from beneath a manhole cover near the water tank, were extracted from the Davis cavern at the time of the measurements described in this paper and uncertainties are given for each individual measurement. \label{tab:radioactivities} }
\begin{ruledtabular}
\begin{tabular}{ l c c c c }
\multirow{2}{*}{\textbf{Sample}}  & & \textbf{$^{40}$K} & \textbf{$^{238}$U} & \textbf{$^{232}$Th} \\
& &  \textbf{(Bq/kg)} & \textbf{(Bq/kg)} & \textbf{(Bq/kg)} \\ \hline 
%\multirow{2}{*}{Homestake} & $297\pm59$ & $2.7\pm0.5$& $1.3\pm0.3$	\\
\multirow{2}{*}{\bf{Homestake}} & \footnotesize avg. & $297$ & $2.7$& $1.3$	\\
 & \footnotesize range & 31--601 & 0.7--9.5 & 1.0--6.5 \\ 
%\multirow{2}{*}{Rhyolite} & $1291\pm258$ & $108\pm22$ & $44 \pm 9$ \\
\multirow{2}{*}{\bf{Rhyolite}} &\footnotesize avg &  $1291$ & $108$ & $44$ \\
& \footnotesize range & 523--2127 & 99--135& 7.7--61 \\
%\multirow{2}{*}{Concrete} & $381\pm76$ &$27\pm5$ & $13\pm3$\\
\multirow{2}{*}{\bf{Concrete}} & \footnotesize avg.&  $381$ &$27$ & $13$\\
 & \footnotesize range & 393--368 &22--27 & 13--14 \\
%Shotcrete (average) & 272.4		&23.3&		11.6	\\
%Shotcrete (1) & 170.3		&18.8&		8.8	\\
%Shotcrete (2) & 380.8		&24.37 &		13.6	\\
%Shotcrete (3) & 244.6		&20.0&		12.5	\\
%Shotcrete & $272\pm54$ & $23\pm5$ & $12\pm2$ \\ 
\multirow{2}{*}{\bf{Shotcrete}} & \footnotesize avg. & $272$ & $23$ & $12$ \\ 
& \footnotesize range & 127--393 & 22-28 & 8.1--14 \\ \hline 
\bf{Shotcrete} & - & $220\pm30$ & $21\pm1$  & $11.4 \pm0.4$\\
\bf{Gravel} & - &  $35.0\pm0.6 $ & $26.3\pm0.1$ & $1.7\pm0.8$  \\ 
\end{tabular}
\end{ruledtabular}
\end{table}

Previous measurements at SURF of the $\gamma$-flux at depths of 800 ft, 2,000 ft, and 4,550 ft demonstrated that the flux at different locations even at the same depth can vary by up to 30\%, depending on the variation in the geological formations~\cite{mei:2009py}. This suggests a direct measurement in the relevant experimental area housing a low-background experiment is required. For the Davis cavern at a depth of 4,850 feet, a measurement with a HPGe detector was taken within a room on the lower level adjacent to the water tank, which prior to this work was the only measurement in the vicinity of the LZ experiment. The $\gamma$-ray flux from this HPGe measurement was stated to be constrained by an upper limit of 2.19~$\gamma$~cm$^{-2}$~s$^{-1}$ (1000--2700~keV), but substantial uncertainties in both the analysis and calibration for this study renders the results too inaccurate for the purposes of background determination for LZ. This motivated a dedicated measurement to obtain a more precise $\gamma$-ray flux, presented in this paper, to be used in LZ background estimations for WIMP search, outer detector background rates, and other relevant rare event searches such as $0\nu\beta\beta$-decay.

%Intrinsic contamination of the liquid scintillator is expected to provide a rate of 60~Hz~\cite{Haselschwardt:2018}

\section{Experimental Setup}
\begin{figure}[H]
\centering
\includegraphics[width=0.65\linewidth]{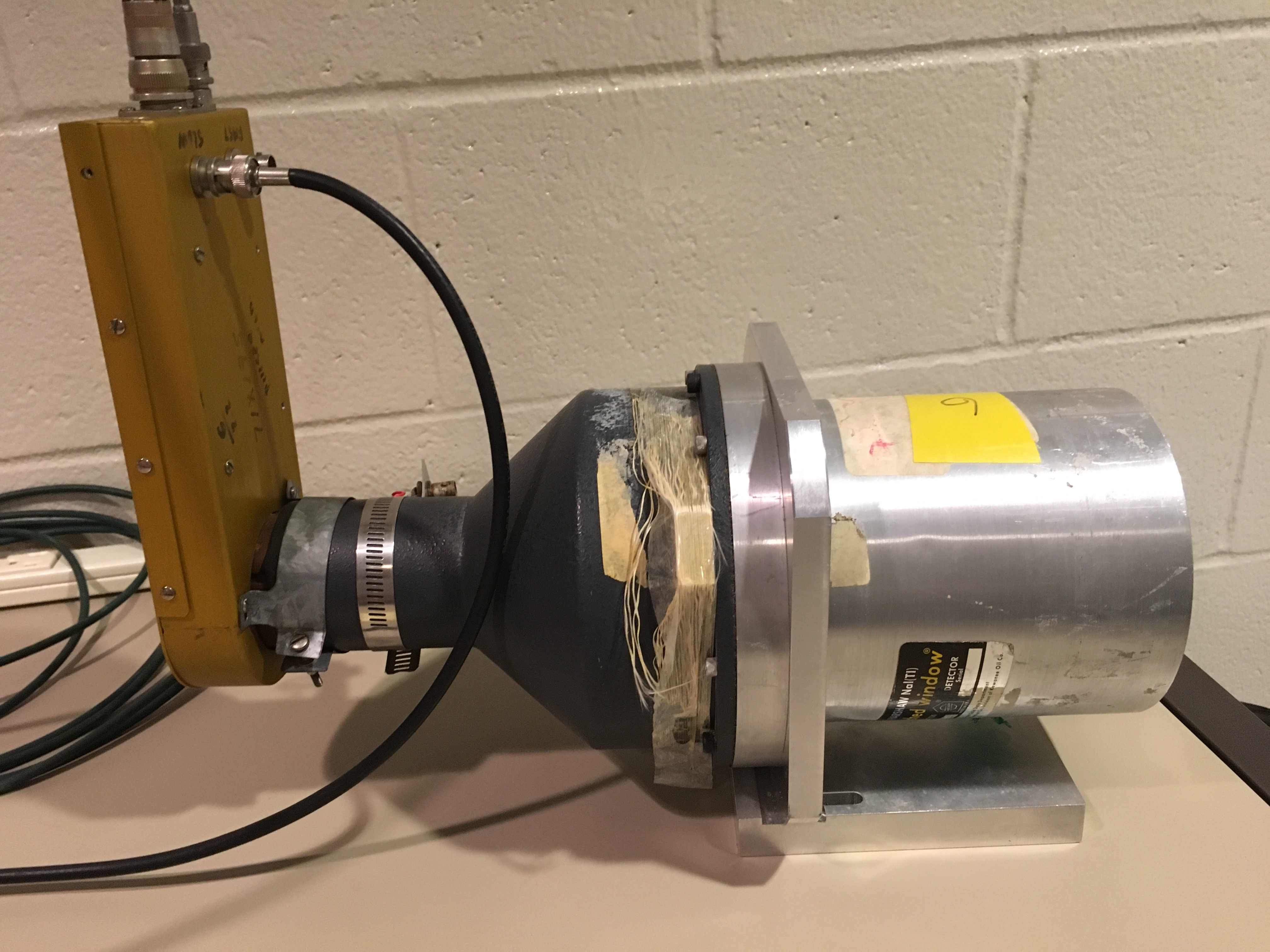}
\caption{ \small A photograph of the 5-inch NaI(Tl) detector, showing the preamplifier, PMT and NaI crystal. \label{fig:NaIDetector}}
\end{figure}

To measure the $\gamma$-ray flux in the Davis cavern, a 5-inch~$\times$~5~inch thallium-doped sodium iodide (NaI) scintillating crystal coupled to a photomultiplier tube (PMT) was used, as shown in Fig.~\ref{fig:NaIDetector}. The detector, manufactured by Harshaw, was connected to a NOMAD 92X-P portable $\gamma$-spectroscopy unit. The MAESTRO software was used to produce spectra \cite{MAESTRO}. 

\begin{figure}[h]
\includegraphics[width=0.49\linewidth]{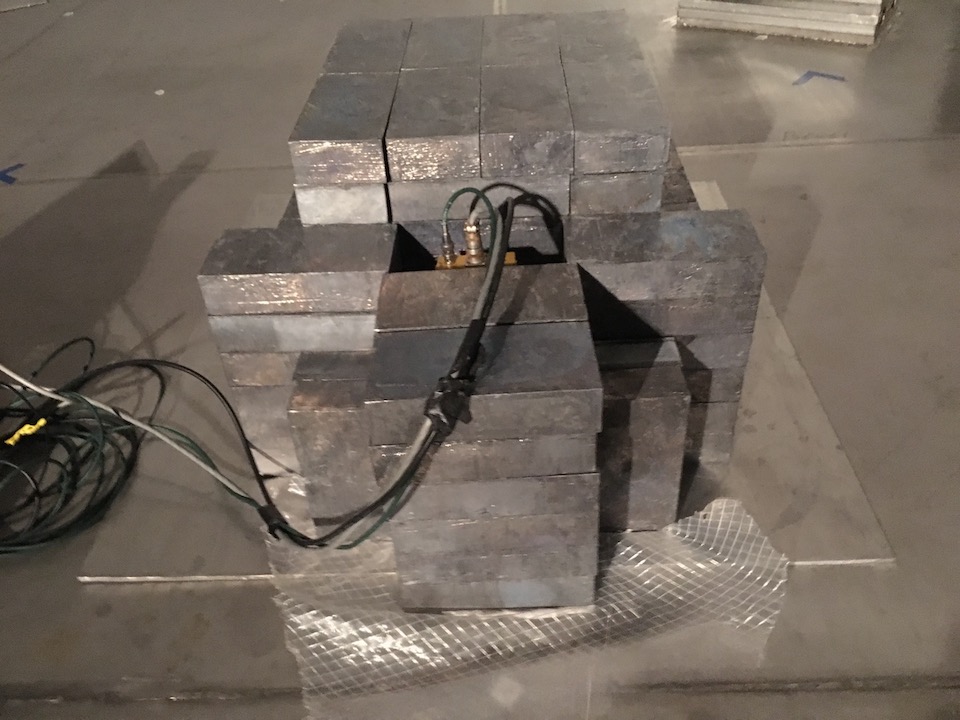}
\includegraphics[width=0.49\linewidth]{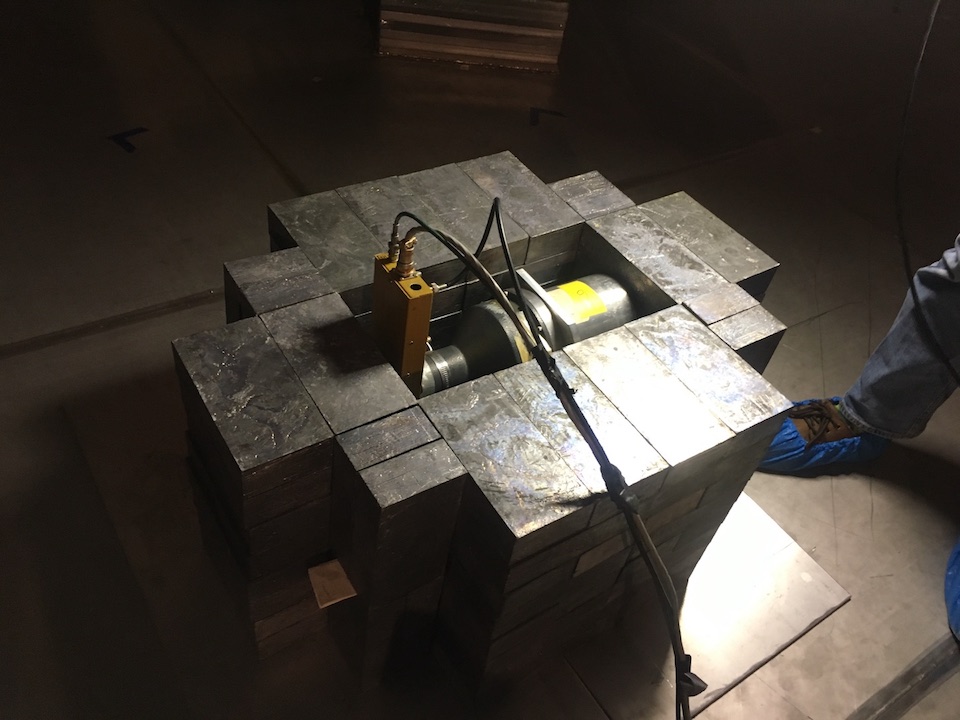}
\caption{Photographs of the lead shields constructed to expose the detector to $\gamma$-rays from below (L) and above (R). \label{fig:lead}}
\end{figure}

A total of 130 lead bricks measuring $8\times4\times2$~inches were used to create $\gamma$-ray shielding for the NaI detector; the lead used was virgin (used direct from ore) and obtained from the Doe Run Mining Company. Three different shielding configurations were used to expose the detector to the $\gamma$-ray flux from different directions within the cavern; the detector was shielded from all sides except below, above, and sideways towards the flat face of the crystal, see Fig. \ref{fig:lead}. This shielding was constructed to provide at least 8~inches of lead on the sides that were not exposed. Many of the measurements both with and without shielding were taken inside the water tank, which had been emptied of water after removal of the LUX detector.

\subsection{Detector Calibration and Efficiency}
%\textcolor{red}{energy vs energy resolution}
%\begin{figure}
%\includegraphics[width=1\linewidth]{newfit}
%\caption{Relative resolution fit}
%\end{figure}

A $^{60}$Co source calibration was performed before each measurement. The gain of the NOMAD unit was adjusted so that the sum peak at 2505~keV was visible, ensuring that the dynamic range was adequate to fully contain the 2614~keV peak from $^{208}$Tl for the background measurements.
$^{60}$Co, $^{137}$Cs and $^{228}$Th calibration sources were used in order to assess detector efficiency and resolution. 
Each calibration spectrum was fitted with Gaussians and exponential backgrounds in order to determine the location and resolution of the 1173~keV, 1332~keV, and 2505~keV peaks for $^{60}$Co,  the 662~keV peak for $^{137}$Cs and the 2614~keV peak from $^{208}$Tl (a product of the $^{228}$Th decay chain). 
All calibrations were performed in the centre of the water tank on the floor; the unshielded measurement in that position was used to subtract the cavern background from the calibration spectra. %These spectra are shown in  Fig.~\ref{fig:calspecs}.
Using the $^{60}$Co data a calibration curve relating the PMT channels to an energy was obtained, and good linearity was observed.
\begin{comment}\begin{figure}[ht]
\centering
\begin{minipage}{0.45\linewidth}
\includegraphics[width=\linewidth]{Figures/co60spec}
\subcaption{}
\end{minipage}
\begin{minipage}{0.45\linewidth}
\includegraphics[width=\linewidth]{Figures/cs137spec}
\subcaption{}
\end{minipage}
\begin{minipage}{0.45\linewidth}
\includegraphics[width=\linewidth]{Figures/Th228spec}
\subcaption{}
\end{minipage}
\caption{Calibration spectra with a Gaussian on an exponential background. The underlying Gaussian curve is shown in red.}
\label{fig:calspecs}
\end{figure}
%Fig of subtracted fits goes here
\end{comment}

%Calibration curves
%\begin{figure}[ht]
%\centering
%\includegraphics[width=1\linewidth]{Figures/Ecal}
%\caption{Energy calibration curve for the NaI detector using MAESTRO software}
%\label{fig:ecal}
%\end{figure}
%From these fits the integrated rate in each peak was calculated. This rate, however includes the efficiency of the detector, which must be extracted and corrected for when comparing the cavern flux measurements to simulation results.

\begin{figure}[h]
\centering
\includegraphics[width=1\linewidth]{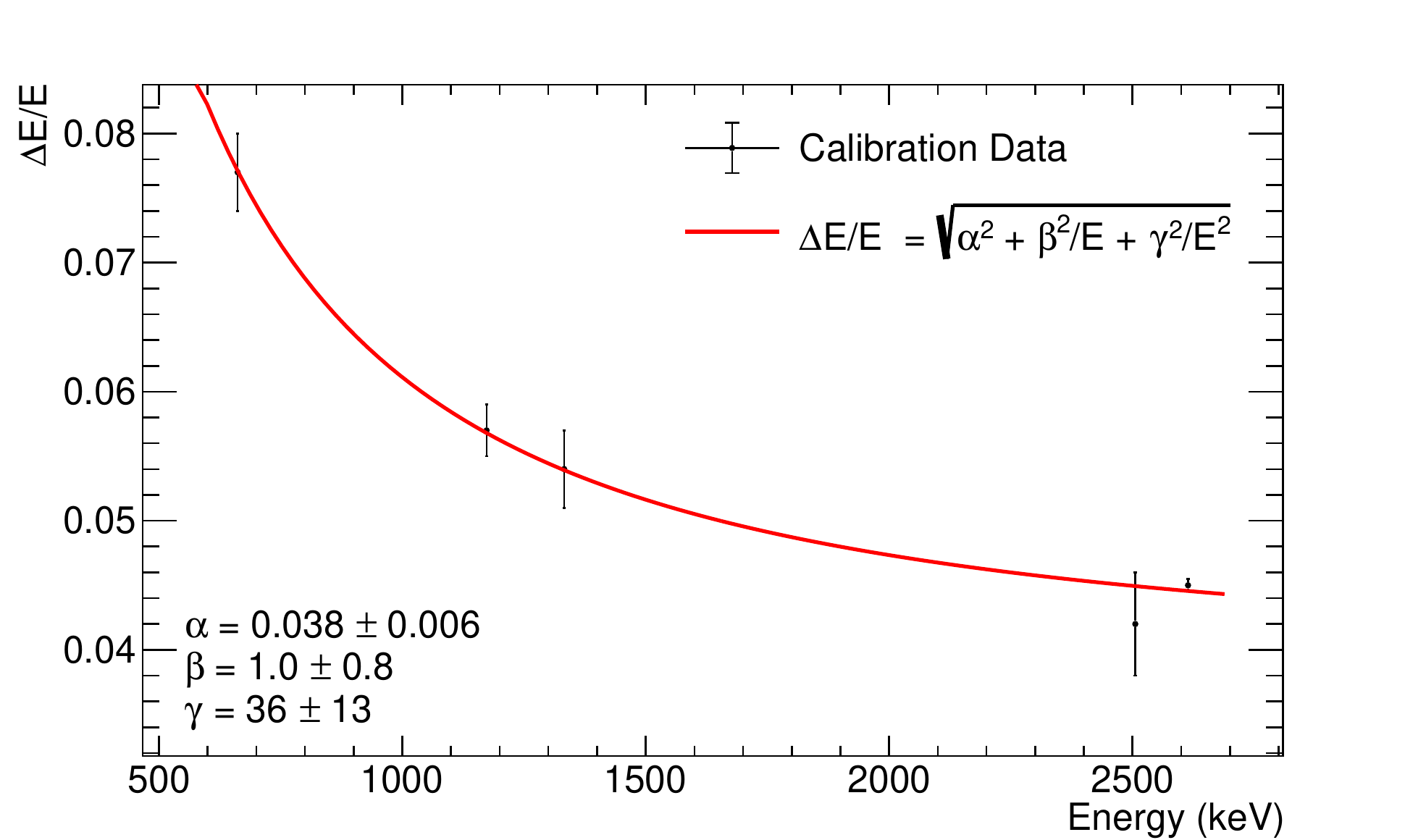}
\caption{\small Resolution of the NaI detector for peaks obtained from $^{137}$Cs, $^{60}$Co and $^{228}$Th calibration data. A fit to the data points using Eq.~\ref{eq:resmodel} is shown. \label{fig:res}}
\label{fig:resfit}
\end{figure}

The absolute efficiency of the detector ($\varepsilon_A$) is a product of the geometric acceptance, which depends on the fraction of solid angle the detector is exposed to, the efficiency describing the conversion of incident $\gamma$-rays in the NaI crystal, and the light collection efficiency of the detector. It can be described as:
\begin{equation}
\varepsilon_A(E)=\frac{N(E)}{A T  P_\gamma(E)},
\label{eq:efficiency}
\end{equation}
where $N(E)$ is the number of counts in a photopeak of energy $E$, $A$ is the activity of the source, $T$ is the live time and $P_\gamma(E)$ is the probability of a single decay producing a $\gamma$ of energy E. 
An over-estimation of $\varepsilon_A$ was observed in simulation; this is most likely an effect of energy-only simulations, as no light collection effects are included. Simulations of calibration sources at varying distances from the detector were performed and source activities were used to calculate the rates in each simulated photopeak. Comparison to data exhibited an average 10\% overestimation of efficiency in simulation with a dependence on distance to the source; a correction factor of $0.90 \pm 0.06$ was therefore applied to further simulations. The detector was also compared to a standard 3~inch~$\times$~3~inch NaI(Tl) crystal as is typical in HPGe spectroscopy; this uses the rate in the 1332.5~keV peak in $^{60}$Co calibration data taken with the source at a distance of 25~cm from the endcap. Note that a percentage greater than 100\% is expected as the detector is 5~inch~$\times$~5~inch and therefore larger in volume by a factor of 4.6. This factor was determined to be 440\% in data, and this was matched by the corrected simulation. 
Furthermore, the resolution $(R)$ of each peak was calculated from the full width at half maximum ($FWHM$) and the energy ($E$) of that peak, using:
\begin{equation}
R=\frac{FWHM}{E} = \frac{\Delta E}{E}.
\label{eq:resolution}
\end{equation}
From these calculations the resolution scale was determined from a fit to the following resolution model~\cite{an:2017}:
\begin{equation}
R=\frac{\Delta E}{E}= \sqrt{\alpha^2+\frac{\beta^2}{E}+\frac{\gamma^2}{E^2}},
\label{eq:resmodel}
\end{equation}
where $\alpha$ describes the light transmission from the scintillating crystal to the photocathode, $\beta$ is the statistical fluctuations in photon production, attenuation, conversion and amplification, and $\gamma$ is the contributions of noise. This resolution model is shown in Fig. \ref{fig:resfit}, and was used to apply a correction to true energy deposits in Monte Carlo simulations of the cavern $\gamma$-flux for direct comparison with the NaI data (see section \ref{sec:simulation}).

\subsection{Data Collection}
\label{sec:datacollection}
\begin{figure*}[ht]
\includegraphics[width=1\linewidth, trim = 50 50 50 50]{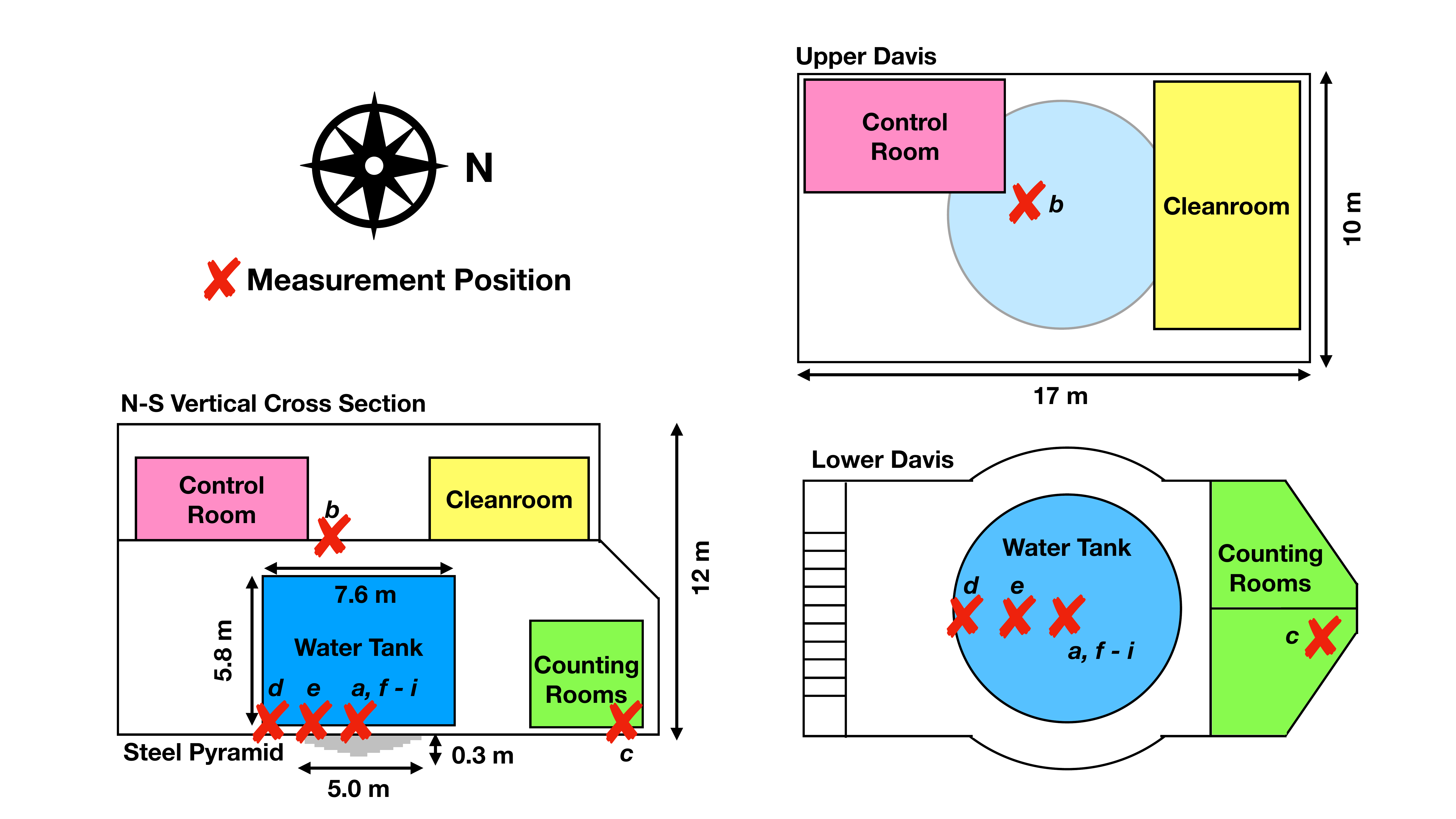}
\caption{ \small Layout of the Davis cavern showing key dimensions and measurement positions denoted with red crosses and labelled as in Table \ref{tab:rates}. Also shown is the location of the LUX/LZ control room, a cleanroom, and the counting rooms. Note that this was the layout at the time of the measurements and changes (such as the removal of the cleanroom) have taken place for LZ construction. \label{fig:DavisLayout}}
\end{figure*}

A total of nine background measurements, three unshielded and six shielded, were taken in the Davis cavern as shown in Fig. \ref{fig:DavisLayout}. Locations for unshielded measurements consisted of on the floor in the centre of the water tank (a), a position on the upper deck in the Davis cavern (known as the Upper Davis) approximately 3.4~m from the water tank centre (b), and within the east counting room (c).
Of the shielded measurements, three were intended to investigate the attenuation due to the steel pyramid beneath the water tank and so were unshielded only from below; the first at the edge of the tank exposing the detector to the gravel beneath (d), positioned at half the radius of the water tank shielded by 15~cm of the pyramid (e), and at the center shielded by the full 30~cm of the pyramid (f).
The final three measurements were with the detector in the centre of the tank; first unshielded from above with the intention of measuring the flux from the ceiling (g), and then the shielding removed from the west facing (h) and east facing sides (i) of the detector at its circular face in order to look for an asymmetry due to the presence of rhyolite.

\begin{figure*}[ht]
\includegraphics[width=1\linewidth]{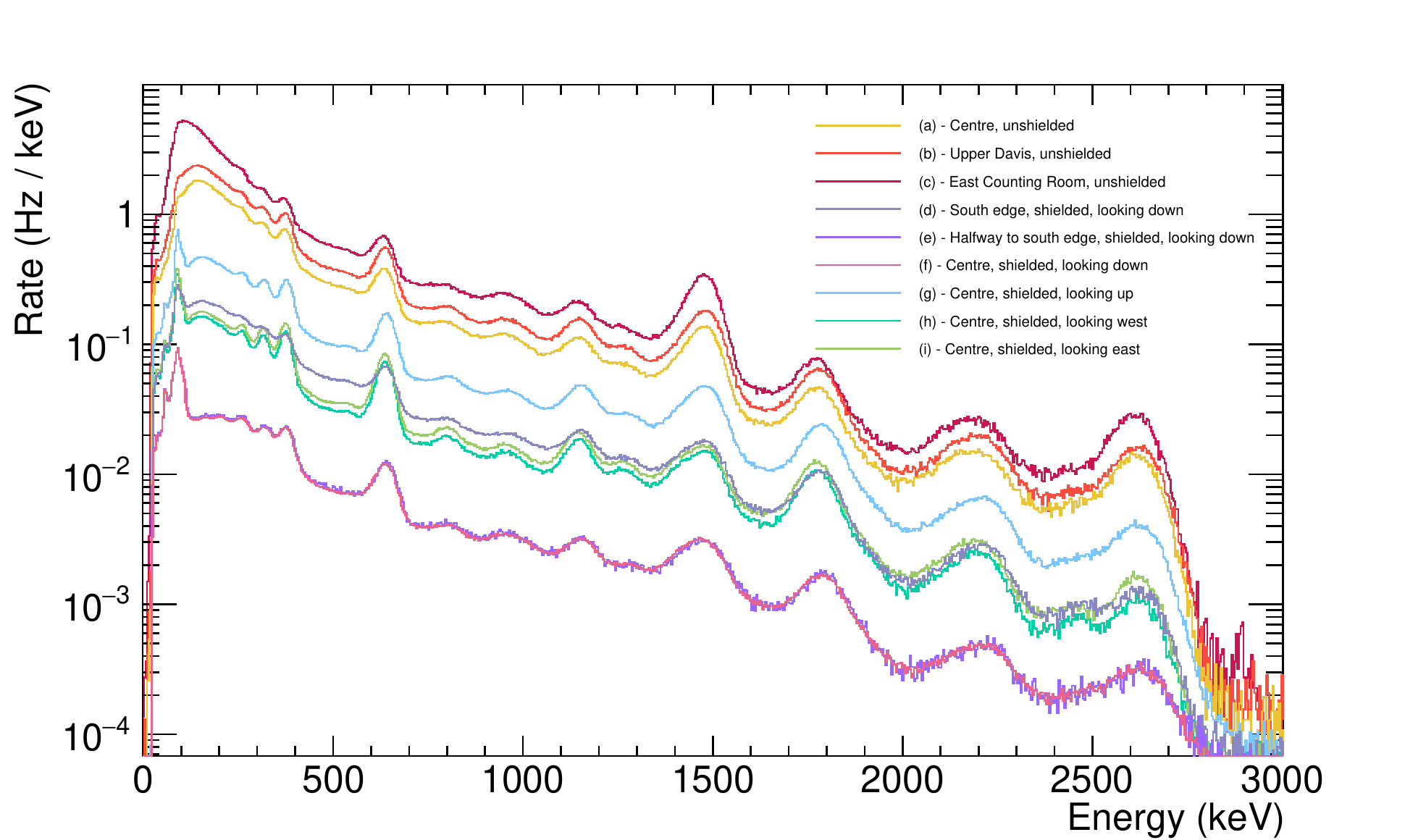}
\caption{\small The energy spectra for all nine measurements in the energy range 0--3000 keV. \label{fig:allspectra} }
\end{figure*}

\begin{comment}

\begin{figure}[h]
\begin{minipage}{0.48\linewidth}
	\includegraphics[width=1\linewidth]{Figures/Upper}
    \subcaption{\footnotesize Upper Davis}
\end{minipage} 
\begin{minipage}{0.48\linewidth}
	\includegraphics[width=1\linewidth]{Figures/CountingRoom}
    \subcaption{\footnotesize East Counting Room}
\end{minipage}
\begin{minipage}{0.48\linewidth}
	\includegraphics[width=1\linewidth]{Figures/Edge}
    \subcaption{\footnotesize  Water tank south edge}
\end{minipage} 
\begin{minipage}{0.48\linewidth}
	\includegraphics[width=1\linewidth]{Figures/LookingDown}
    \subcaption{\footnotesize Downwards view shielding}
\end{minipage}
\begin{minipage}{0.48\linewidth}
	\includegraphics[width=1\linewidth]{Figures/LookingUp}
    \subcaption{ \footnotesize Upwards view shielding }
\end{minipage}
\begin{minipage}{0.48\linewidth}
	\includegraphics[width=1\linewidth]{Figures/WestEast}
    \subcaption{\footnotesize  West/east view shielding}
\end{minipage}

\caption{ \small  Photographs of the NaI detector in  six of the measurement positions and lead shielding configurations.}
\end{figure}
\end{comment}
The energy spectra up to 3000~keV for all 9 positions are shown in Fig. \ref{fig:allspectra}.
Table \ref{tab:rates} contains the total integrated rates and rates integrated from 200--3300 keV. 
The differing live times were dictated by time available underground at SURF, with shielded measurements given priority for overnight data-taking to account for the lower rate.
The overall measured rates were highest in the east counting room, followed by the upper level of the Davis cavern and the centre of the water tank. Differences in rate can be attributed to shielding, the proximity to the steel pyramid and differences in the structure and material of the floor.

For the west and east facing measurements, no significant asymmetry was observed in comparison to the differences in activity of the rhyolite versus the Homestake rock and shotcrete. The rhyolite intrusion is thought to be present on the west wall (see Fig. \ref{fig:rhyolite}), but the rate from this direction was measured to be about 10\% less than the east facing measurement in total rate. This suggests a lack of a significant flux of $\gamma$-rays from rhyolite surviving through the shotcrete layer, and the observed difference may be due to unevenness in shotcrete thickness since the shotcrete is approximately 10 times more radioactive than the Homestake formation in $^{238}$U and $^{232}$Th. 
%To assess the contribution from the floor, the measurement at the centre of the tank can be subtracted from the upper Davis measurement to give 198~Hz. When compared to the measurement looking upwards at the ceiling, 204~Hz, this again suggests a similar flux from above and below. 

For the measurements facing downwards at the steel shielding pyramid, the total rate measured above 30~cm of steel in the centre was within 3\% of the rate in the position at half of the tank radius, where the detector was just shielded by 15~cm of steel, as can be seen by the two lowest rate histograms overlapping in Fig. \ref{fig:allspectra}. The measurement at the edge of the tank, with no steel beneath, had a higher rate. As the gravel under the water tank is known to be relatively low in radioactivity, we assume that the rate measured in (e) and (f) - the lowest - correspond to the intrinsic background of the experimental setup, including the NaI detector, as NaI(Tl) crystals are known to have intrinsic $^{40}$K, $^{238}$U and $^{232}$Th contamination \cite{Adhikari:2016} and PMTs are also known sources of radioactivity. Contributions from the lead shielding are expected to be subdominant to the detector.

Radon in the cavern air resulting from natural emanation must also be considered; decays from below $^{222}$Rn make up a majority of the $\gamma$-rays in the $^{238}$U chain. There is no specific mitigation of underground radon levels at SURF, where the total average radon concentration in the Davis Campus is approximately 310~Bq/m$^3$, with a seasonal dependence resulting in a winter low of 150~Bq/m$^3$~\cite{heise:2015}. However, during some days of data acquisition, unusually high radon levels were recorded, due to changes in airflow in the mine drift outside of the Davis Campus. The average recorded radon concentration (measured with an AlphaGuard detector) for each dataset is also shown in Table~\ref{tab:rates}, but it should be noted that radon data was taken in an area outside the main entrance to the cavern known as the common corridor, so some uncertainties remain over the air circulation differences between this location and where the $\gamma$-ray data was taken, but the concentrations were significant enough that $\gamma$-rays from radon must be included in this analysis. 

\begin{table*}[ht]
\caption{\small Measurement dates, live times, radon concentrations and integrated count rates.  Here, `looking down (up)' refers to the shielding configuration where only the underside (topside) of the detector is not shielded by lead. Uncertainties on rates are Poisson counting errors only. \label{tab:rates}}
\begin{ruledtabular}
\begin{tabular}{ l c c c c c c} 
\multirow{2}{*}{\textbf{Position of measurement}}  &\multirow{2}{*}{\textbf{Label}} & \multirow{2}{*}{\textbf{Start Date}} &  \textbf{live time} &  \textbf{Avg. Radon} & \textbf{Rate (Hz)} & \textbf{Rate (Hz)}  \\
& & & \textbf{(hours)} & \textbf{(Bq/m$^3$)} & \textbf{Total}  & \textbf{$> 200$ keV}  \\ \hline 
Centre of water tank, unshielded & a & 24/10/17 & 4.0  & $422 \pm 34$& $595.7\pm 0.2$  & $386.0\pm0.2$ 	\\
Upper Davis, unshielded & b & 26/10/17 & 3.6 & $868 \pm 222$& $794.4\pm0.2$   &  $512.0\pm0.2$\\
East Counting Room, unshielded &c & 26/10/17&  2.1 & $929 \pm 70$& $1355.0\pm0.4$ & $750.9\pm0.3$  \\ 
Edge of tank, looking down &d & 16/10/17& 18.2 & $358\pm80$ &$94.17\pm0.04$ &\ $64.40\pm0.03$ \\
Halfway to edge of tank, looking down  &e &  17/10/17 & 17.9 & $336 \pm 55$ & $17.15\pm0.02$ & $10.70\pm0.01$ \\
Centre of tank, looking down & f &19/10/17& 117.0 & $500 \pm 155$ & $16.715\pm0.006$ & $10.427\pm0.005$  \\ 
Centre of tank, looking up &g &18/10/17 & 20.2 & $372 \pm 76$ & $203.57\pm 0.05$ &  $139.0 \pm 0.04$  \\ 
Centre of tank, looking west &h & 24/10/17 & 17.3 & $359\pm37$ & $95.11\pm0.04$  & $51.77\pm0.03$\\
Centre of tank, looking east &i & 25/10/17 & 22.3 & $316\pm 46$& $106.33\pm0.4$ &  $59.14\pm0.03$\\
\end{tabular}
\end{ruledtabular}
\end{table*}

\section{Simulation and Analysis}
\label{sec:simulation}
Simulations of the cavern background are performed using the BACCARAT framework, a \textsc{Geant4} v.10.03~\cite{Agostinelli:2002hh} package primarily used for LZ background simulations. The model used for electromagnetic processes is \textit{G4EMLivermorePhysics}; this covers interactions using Livermore models for $\gamma$ and electron cross-sections \cite{liv1} \cite{liv2}, with a  focus on low energy processes, such as Rayleigh and Compton scattering, bremsstrahlung and the photoelectric effect.
An event biasing technique is applied to accelerate the simulation; this was developed for LZ background simulations of the cavern due to the low probability of a $\gamma$-ray surviving through the water shield, outer detector and the skin layer of liquid xenon. 
$^{238}$U, $^{232}$Th and $^{40}$K decays are initiated within a 30~cm thick layer of material surrounding the cavern at the approximate location of the cavern walls. The event biasing technique involves saving $\gamma$-rays  on a predefined surface, then propagating them onward with the same momentum in a second simulation with a multiplicative factor in order to increase $\gamma$-ray statistics. This can be done multiple times on surfaces of decreasing size around a target detector volume. 
%The integrated fluxes from ref. \cite{mei:2009py} for the three principal lines - 1,460~keV for $^{40}$K, 1,764~keV from $^{214}$Bi for $^{238}$U  and 2,614~keV from $^{208}$Tl for $^{232}$Th - were used to obtain rock activities that in simulation resulted in the correct fluxes in the cavern through a surface of dimensions $7.82\times7.82\times6.68$~m. These activities were 716~Bq/kg of $^{40}$K, 73.4~Bq/kg of $^{238}$U and 26.1~Bq/kg $^{232}$Th. 
%These were applied to LZ simulations using the event biasing technique and were found to result in an event rate in the LZ outer detector of O(100 Hz). The total rate inside the outer detector including from LZ components and intrinsic contamination within the scintillator is constrained to be $<100$~Hz, in order to impact $<5$\% of LZ live time whilst utilising a 500$\mu$s veto window. The current prediction 

A custom geometry featuring the cavern, pyramid, water tank and detector was created, with options to use each of the lead shielding configurations to expose the NaI detector from either above or below. The cavern and the surrounding rock are modelled as a cuboid with internal space of $20\times14\times12$~m; this is larger by design than the dimensions shown in Fig. \ref{fig:DavisLayout} to conserve the simulated surface area with reality, as the cavern walls are uneven. Radioactive sources were placed within a 30~cm thick `shell' on the inside of this cavern rock. A simulation study done for experiments at the Modane Underground Laboratory have shown that due to attenuation, a 30~cm concrete/shotcrete shell configuration is sufficient to produce more than $96\%$ of the total $\gamma$-ray flux without needing to include emission from the underlying rock~\cite{tomasello:2010zz}. This approximation is expected to hold also for the Davis cavern; in places, both the wall and the concrete are 30~cm thick, and the ratio of shotcrete to rock activities is in most cases higher than at the Modane Underground Laboratory, making the rock contribution less prominent.  The cavern rock is simulated using the chemical composition of a Homestake sample and is a mixture of oxides, primarily SiO$_2$, Al$_2$O$_3$, FeO and water~\cite{mei:2009py}. Discrepancies between simulation and data for different measurement positions may indicate a deviation from this simplified model, such as a contribution from $\gamma$-rays from the more radioactive rhyolite beneath the shotcrete. 

\begin{figure}[h]
\includegraphics[width=1\linewidth]{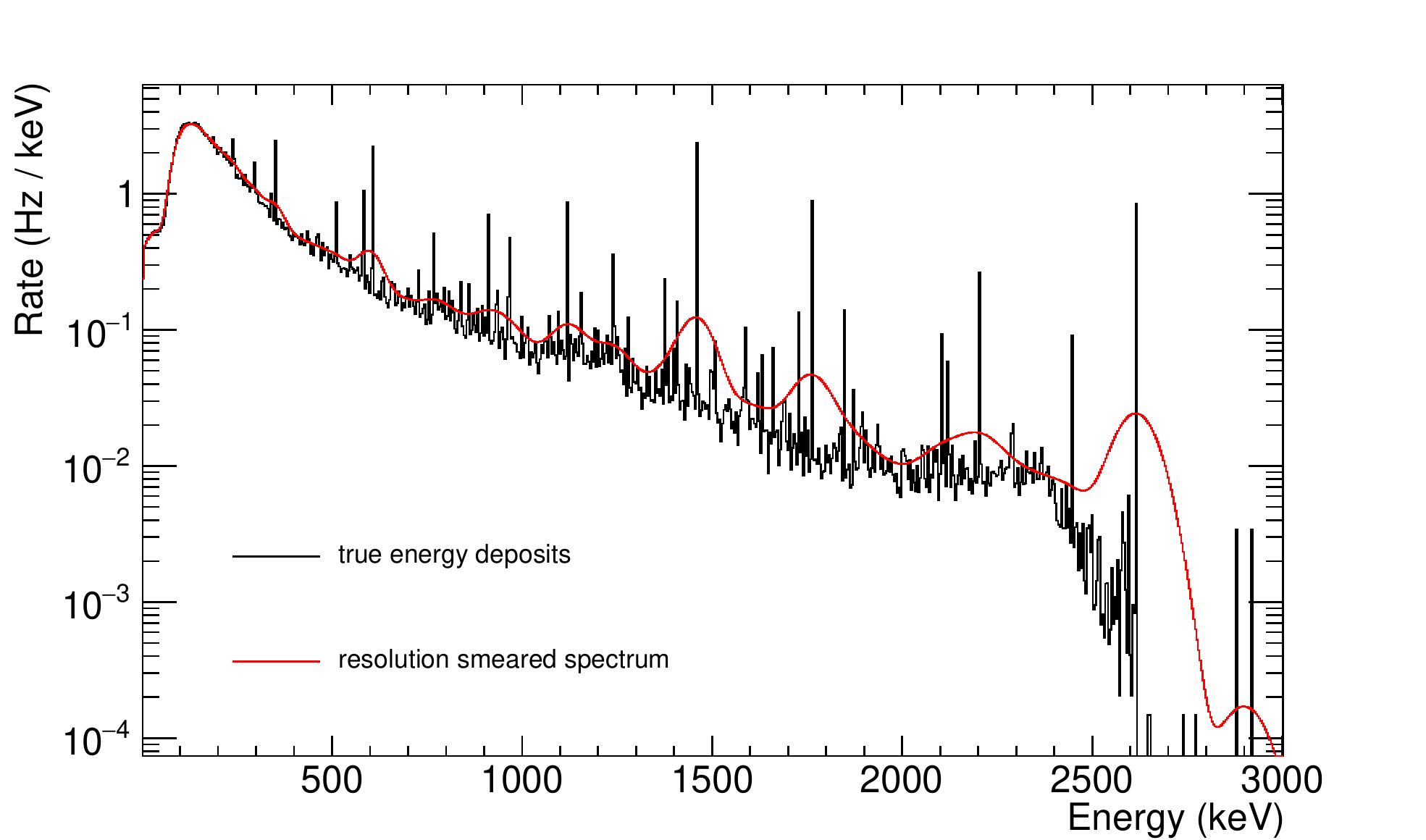}
\caption{\small An example energy spectrum obtained in simulation for position (a). The black histogram shows true energy deposits where photopeaks from various lines are visible, whilst the red line shows the result of smearing with Eq. \ref{eq:resmodel}. \label{fig:smearedsim}}
\end{figure}

For the $^{238}$U and $^{232}$Th chains, an event generator developed for LZ background simulations was used. For each event, this generator initiates a chain of decays beginning at $^{238}$U or $^{232}$Th and ending at the stable $^{206}$Pb or $^{208}$Pb. Therefore, $\alpha$, $\beta$ and $\gamma$-decays for the entire chain are generated with the correct energies and branching ratios. Secular equilibrium is assumed, as any break in equilibrium is not expected to have an effect on the determined concentration of radioactive contaminants, since the high-energy and high-intensity $\gamma$-lines measured in this analysis are all from from the late sub-chains.
%due to the most prominent $\gamma$-lines existing in the late chains.
%The validity of the simulation was confirmed by comparing relative rates for the different measurement positions to the data. For example, the two positions where the rate was lowest recorded the same rate in the simulation. Additionally, the highest rate measurement in the east counting room was found to have a total rate 2$\times$ bigger than the rate in the centre of the water tank in both simulation and data, despite the lack of features such as the counting room walls and the true shape of the cavern. 
Energy deposits by conversion of $\gamma$-rays inside in the NaI crystal were recorded and then smeared by a Gaussian function using parameters from the fit of Eq.~\ref{eq:resmodel} to calibration data shown in Fig.~\ref{fig:resfit}. This was done separately for $^{40}$K, $^{238}$U and $^{232}$Th.  
Fig.~\ref{fig:smearedsim} shows an example of a simulated energy spectrum summed over $^{40}$K, $^{238}$U and $^{232}$Th in true energy deposits before and after smearing. 

Additionally, due the high levels of radon present at the time of the measurements, the $^{222}$Rn decay chain was simulated within the cavern air and inside the water tank and the rates normalised using the measured concentrations. 

%There are some notable discrepancies between the three unshielded measurements. In particular, whilst $^{40}$K and $^{232}$Th are in reasonable agreement between the centre of tank and east counting room measurements, the $^{238}$U differs by a factor of \textcolor{red}{1.6}. Both the measurement on the upper level of the Davis and the `looking up' measurement inside the water team seem to suggest a lower concentration of  $^{40}$K. The rhyolite intrusion is known to contain a significant amount of potassium - 1291~mBq/kg, and as this is known to pass beneath the cavern but not above it, may explain the smaller data to simulation ratio for these two measurements.

For use in future LZ simulations, the simulated activities of isotopes within the shell in Bq/kg are required to reproduce the measured rate in the NaI data. Since the Compton background in the simulation is unlikely to be accurate due to the lack of many features of the real cavern and objects within it in the simulation, the analysis focuses on $\gamma$-lines with energies of 1400~keV and above where the Compton background is less dominant. The simplest technique is to fit Gaussians for the four most prominent lines at 1461~keV ($^{40}$K, BR: 10.66\%) ), 1764~keV ($^{214}$Bi, BR: 15.30\%) , 2204~keV ($^{214}$Bi, BR: 4.92\%) and 2614~keV ($^{208}$Tl, BR: 99.75\%). Choosing to focus on the photopeaks selects predominantly $\gamma$-rays that have been produced near the surface of the cavern walls, as they have travelled to the detector without any Compton scatters.

\begin{figure}[h]
\centering
\begin{minipage}{1\linewidth}
\includegraphics[width=\linewidth, trim = 10 10 10 10]{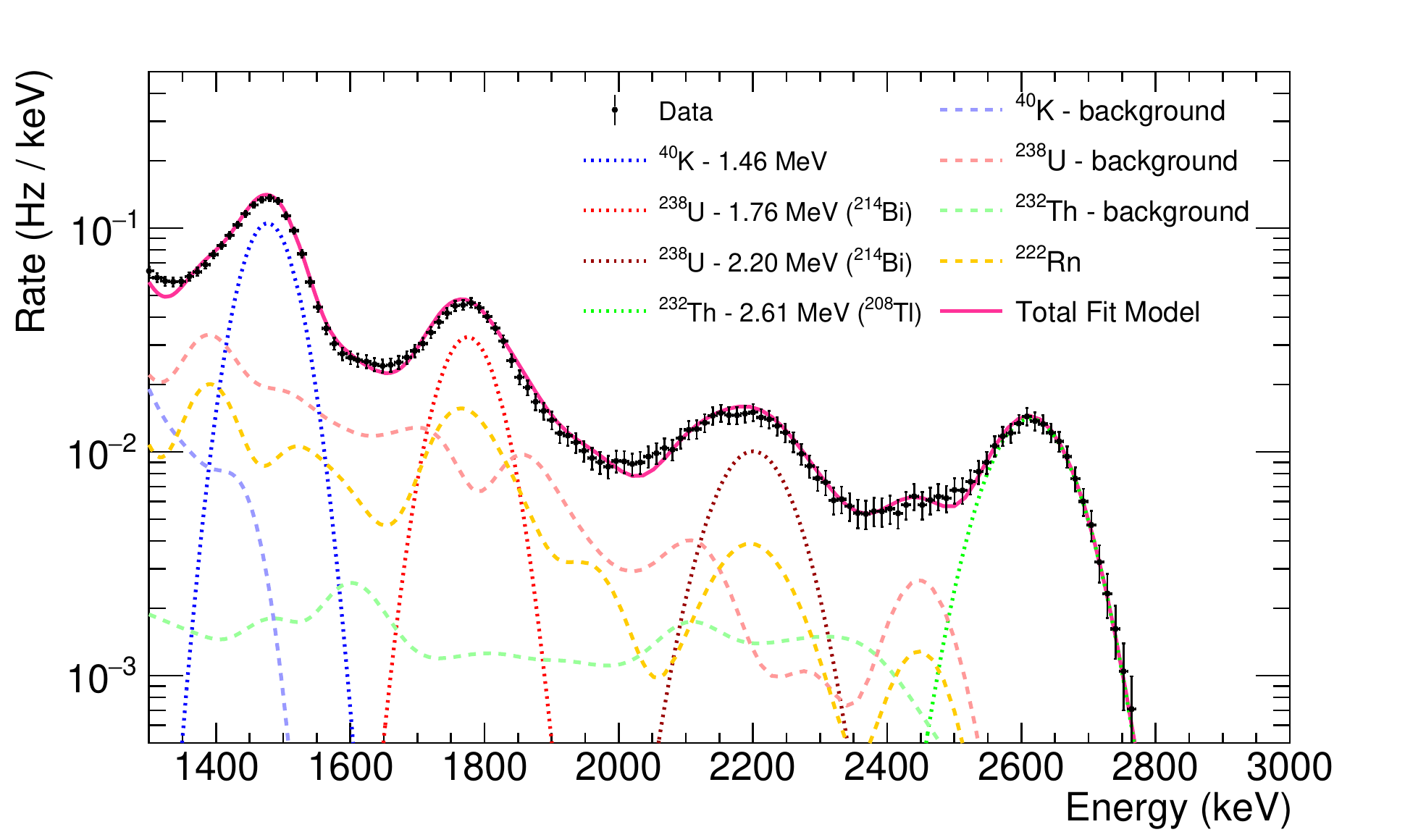}
\end{minipage}
%\begin{minipage}{0.9\linewidth}
%\includegraphics[width=\linewidth]{Figures/pos1}
%\end{minipage}
%\begin{minipage}{0.9\linewidth}
%\includegraphics[width=\linewidth]{Figures/pos5}
%\end{minipage}
\caption{\small Fitted energy spectrum for position (a) showing the  1461~keV $^{40}$K line, the 1764~keV and 2204~keV lines from $^{238}$U and 2614~keV from $^{232}$Th, background contributions from other, less dominant lines and Compton scattering, and in yellow, the airborne radon contribution.}
\label{fig:goodfits}
\end{figure}

The background shape was difficult to model, as it contains Compton scatters from $\gamma$-rays of all energies and other peaks from the uranium and thorium chains. A background probability distribution function (PDF) was created from the simulated spectra with the most prominent lines removed, and signal PDFs were produced from Gaussian functions with widths constrained within error bars obtained from the resolution function shown in Fig. \ref{fig:resfit}. Several constraints were placed on the fit; the peak to continuum ratio was constrained to float within 20\% of the simulated value for each contribution, the branching ratio for the two $^{214}$Bi lines was fixed, and the total rate from radon was allowed to float with a 20\% uncertainty due the location of the radon measurement.  %This allowed the mean and width of the peaks to float. For the 2.20~MeV line, a superposition of lines at 2.12 and 2.20~MeV at branching ratios of 1\% and 5\% respectively led to the use of a skew Gaussian to obtain the best fit. 
An example of the fit to the unshielded data taken in the centre of the water tank is shown in Fig. \ref{fig:goodfits}. The energy region below 1300~keV is not shown; some non-linearity of the channel to energy calibration and the lack of resolution measurements below 662~keV for smearing of the simulation leads to poorer agreement. 
For each of the 1461, 1764 and 2614~keV peaks (denoted by the subscript $i$ for isotope) in each measurement (denoted by the subscript $m$),  a corresponding activity was determined using the comparison to the simulated rate in each peak:
\begin{equation}
A_{i,m} = \frac{R_{i,m}-R^I_i}{\varepsilon R^{sim}_{i,m}},
\end{equation}
where $R_{i,m}$ is the rate in each signal peak in data determined by the Gaussian fit, $R^I_m$ is an internal rate calculated from the average of measurements (e) and (f) - these were shielded by both lead and the steel pyramid so are subtracted to account for the intrinsic background of the set up, $\varepsilon$ is the aforementioned efficiency correction determined from calibration, and the rate in the simulated peak, $R^{sim}_{i,m}$ is calculated using:
\begin{equation}
R^{sim}_{i,m} = \frac{N_{i,m}}{N^{tot}_{i,m}  B_{m}  M},
\end{equation}
where  $N_{i,m}$ is the raw number of counts in a given peak for isotope $i$ and measurement position $m$, $N^{tot}_{i,m}$ is the total number of initial events simulated, $B_{m}$ is the event biasing multiplicative factor for measurement $m$, and $M$ is the mass of the simulated shell. This is equivalent to simulating 1 Bq/kg for each isotope and determining the necessary scale factors to match data. 
\section{Results and Discussion}
Table \ref{tab:peakrates} contains the fit results for each peak in both rate, $R_{i,m}$ and activity $A_{i,m}$. There are large variations in the activities determined for each position; this is a result of several unknown factors. Firstly, the radon concentration within the cavern, whilst allowed to float within 20\% of the measurement in a separate area of the Davis campus, produced a very similar spectrum to the uranium from the cavern walls, since most $\gamma$-rays are emitted from the late chain decays after $^{222}$Rn. Due to a lack of detailed information about airflow around the Davis campus at this time and the unusual fluctuations in radon concentration, however, this uncertainty is expected to account for the large variation in $^{238}$U activities.
\begin{table*}[ht]
\caption{\small Fit results for the three signature lines for each isotope/decay chain. The best fit activities, $A_m$, are given for each measurement - except the two lowest rate positions, where the contribution from the cavern is minimal. The uncertainties are from fit results only; a larger systematic uncertainty can be expected from the simplified simulation model. The average values are show at the bottom with their standard deviations. \label{tab:peakrates}}
\begin{ruledtabular}
\begin{tabular}{l c |c c | c c | c c} 
\footnotesize
\multirow{3}{*}{\textbf{Position of measurement}} & \multirow{3}{*}{\textbf{Label}} & \multicolumn{2}{c|}{\textbf{$^{40}$K - 1461 keV}} & \multicolumn{2}{c|}{\textbf{$^{238}$U - 1764 keV}}  &\multicolumn{2}{c}{\textbf{$^{232}$ Th - 2614 keV}}\\
& & \footnotesize Rate & \footnotesize $A_{m}$ & \footnotesize Rate & \footnotesize  $A_{m}$ & \footnotesize Rate & \footnotesize $A_{m}$ \\ 
 & & \footnotesize(Hz) &\footnotesize (Bq/kg) & \footnotesize(Hz) & \footnotesize(Bq/kg) & \footnotesize(Hz) & \footnotesize(Bq/kg) \\
\hline
Centre of water tank, unshielded& a & $10.33\pm0.04$ &$285\pm1$ &$2.55\pm0.02$ &$36.9\pm0.4$ & $2.12\pm0.02$ & $15.2\pm0.1$\\
Upper Davis, unshielded& b & $13.82\pm0.07$  & $135\pm4$  &$1.56\pm0.02$ &$10.4\pm0.2$ & $2.498\pm0.004$ & $8.8\pm0.1$ \\
East Counting Room, unshielded&c  &$28.8\pm 0.1$ & $264\pm1$   & $2.98\pm0.03$ & $18\pm 0.2$ & $4.31\pm0.03$ & $12.2\pm0.2$ \\ 
Edge of tank, looking down  & d & $1.01\pm 0.01$& $182\pm2$  & $0.875\pm 0.005$  &$31.4\pm0.2$ & $0.214\pm 0.002$ & $16.7\pm0.1$\\
Halfway to edge of tank, looking down & e  & $0.167 \pm  0.002$  &-   & $0.141\pm 0.002$ & -& $0.0487\pm0.0001$  &-\\
Centre looking down  & f & $0.300\pm0.001$ & -  & $0.157\pm 0.001$ & - & $0.0560\pm0.0004$ & - \\
Centre of tank, looking up &g & $3.47\pm0.01$  & $214\pm1$ & $1.29\pm0.01$ & $48.4\pm0.2$  & $0.650 \pm 0.004$ & $9.5\pm0.1$\\ \hline 
Averaged activities & - &- & $220 \pm 60$ & -& $29 \pm 15$& - & $13 \pm 3$\\
\end{tabular}
\end{ruledtabular}
\end{table*}
Furthermore, the simulation uses a simplified cavern geometry lacking many features of the cavern, such as steel grating, the walls of the counting rooms and the control room, which may affect the peak to continuum ratios. 
Finally, a variation in activities for different measurements may indicate some non-uniformity in the concentrations of each isotope spatially within the cavern walls, although not high enough to suggest the presence of rhyolite. A range of activities were measured for each isotope in Table~\ref{tab:radioactivities}, and depending on the source of the shotcrete material, the values of these may have a spatial dependence.

The treatment of each measurement as an independent observation of the same flux within the cavern results in average activities of $220\pm60$~Bq/kg of $^{40}$K, $29\pm15$~Bq/kg of $^{238}$U and $13 \pm 3 $~Bq/kg of $^{232}$Th. The total measured $\gamma$-ray flux above 0~keV is $1.9 \pm 0.4$~$\gamma$~cm$^{-2}$~s$^{-1}$ and for energies exceeding 1000~keV is $0.35 \pm 0.08$~$\gamma$~cm$^{-2}$~s$^{-1}$, as show in Table~\ref{tab:fluxes}. 
This is consistent with the upper limit of 2.19~$\gamma$~cm$^{-2}$s$^{-1}$ above 1000~keV in Ref.~\cite{thomas:2014}. An additional flux of $0.09\pm0.02$~$\gamma$~cm$^{-2}$~s$^{-1}$ can be expected from the yearly average radon activity of 310~Bq/m$^3$ in normal ventilation conditions. Normalising simulated rates in each peak with these average activities gives good agreement across all measurements and is considered sufficient to estimate background rates for the LZ experiment. If necessary, this flux can be scaled with the radon levels measured at SURF, if more large excursions are observed. 

A quick cross-check between simulation and data was performed using the prominent 609~keV $^{214}$Bi line (BR: 45.5\%). Due to the non-linearity in energy calibration, this appeared at 635~keV in the data, but the rate within the peak could be compared to a simulation normalised using the averaged activities and constrained radon rate. For example, in position a, the rate within the 609~keV peak was measured to be $10.6\pm0.6$ Hz, whilst simulation produced $10\pm3$ Hz. Furthermore, this is consistent with the relative branching fraction ratio of 609~keV to 1764~keV of 2.97, as the total (combined $^{238}$U in the walls and $^{220}$Rn in air) rate within the 1764 keV peak was $3.59\pm0.03$~Hz. 

\begin{table}[h]
\caption{\small Integrated $\gamma$-fluxes in the Davis cavern from radioactive contamination within the walls. \label{tab:fluxes}}
\begin{tabular}{c  c}  \hline \hline 
\textbf{Energy} & \textbf{Flux} \\ 
\textbf{(keV)} & \textbf{($\gamma$ cm$^{-2}$ s$^{-1}$)}\\ \hline  
0--1000 & $1.6 \pm 0.4$  \\
1000--2000  & $0.30 \pm 0.08 $ \\
$>2000$ & $0.05 \pm 0.01$ \\ \hline 
Total & $1.9 \pm 0.4$ \\
\hline \hline 
\end{tabular}
\vspace{-10pt}
\end{table}

Shotcrete undergoing HPGe screening contained $^{40}$K,  $^{238}$U, $^{232}$Th levels of 272~Bq/kg, 23~Bq/kg and 12~Bq/kg respectively for averages of samples during construction, and the more recent sample contained $220\pm30$~Bq/kg, $21\pm 1$~Bq/kg, $11.4\pm0.4$~Bq/kg~(see Table \ref{tab:radioactivities}), showing agreement within uncertainties with the results of this analysis. This suggests the dominant contribution to the $\gamma$-ray flux in the Davis cavern is the shotcrete layer, with no measurable excess or directionality due to rhyolite. 
%Steel samples taken from the water tank and safety grating during construction of LZ were screened with HPGe detectors to establish whether a contribution from this could also be ignored. The steel had uranium, thorium and potassium contamination levels all below 10~mBq/kg and so are also considered negligible compared to the cavern.

%\begin{table}
%\caption{Estimated $\gamma$-flux at the water tank surface}
%%\begin{tabular}{c|l|l|l}
%Isotope & $^{40}$K & $^{232}$Th & $^{238}$U \\
%\end{tabular}
%\end{table}

\section{Comparison with the LS Screener and Implications for LZ}
\begin{figure}[h]
\includegraphics[width=1\linewidth]{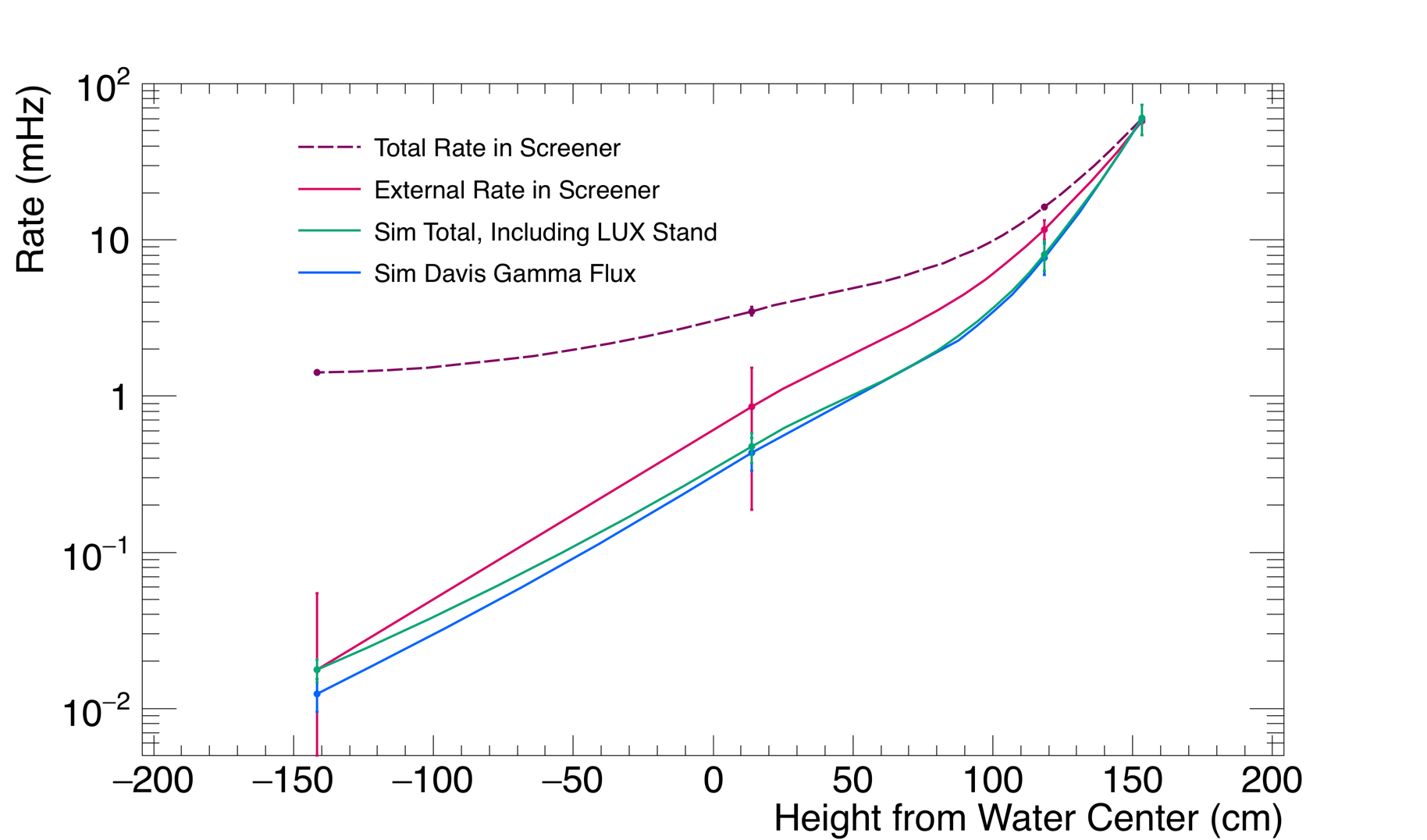}
\caption{\small Event rates above 1300~keV within the filled water tank at four positions in z, as measured by the LS Screener detector (dashed purple). A subtraction of the internal rate was performed and the corresponding external rate is shown as solid magenta, for comparison with the simulated cavern rate.  Note that an additional subtraction with a large uncertainty has been performed on the external rate data point just above 0~cm due to an uncertain level of radon from mine air dissolved in the water after the water tank was opened to move the LS Screener between measurements. \label{fig:screener} }
\end{figure}

A cross-check of these results was performed through simulation of the LS Screener detector, described in Ref.~\cite{Haselschwardt:2018}, which took data at four positions in $z$ within the water tank. The $\gamma$-ray contribution to the rate in the LS Screener from both the cavern and a steel stand originally used to support the LUX detector was simulated within the BACCARAT simulation framework using the event biasing technique. Due to the attenuation of cavern $\gamma$-rays by the water in the tank, initial simulated effective decays (including multiplicative factors) totalled $1\times10^{14}$ per isotope, and two event biasing surfaces were used to obtain sufficient statistics within the detector. These simulation results were combined with the internal rate of the LS Screener measured in the position best shielded from external $\gamma$-rays ($\sim140$~cm below the centre of the tank), see Fig.~\ref{fig:screener}. In the position most sensitive to the cavern flux at the top of the water tank, the rate above 1300 keV predicted by simulation using the activities determined in this paper was $60 \pm 10$~mHz, consistent with with the observed rate of $60 \pm  1$~mHz. 

The LZ Outer Detector is designed to operate with a veto threshold of 200~keV; above this, the integrated rate expected from the cavern using the results of this analysis is $27\pm7$~Hz.
%To validate this, we compare total rates above 200 keV in simulation and data for both the NaI measurements and the LS Screener and find an average excess in simulation of a factor of $2.0\pm0.2$. This suggests a lower peak to continuum ratio within both simulations. This is somewhat expected due to the simple geometrical modelling of the cavern; in reality, there is far more material in the form of structures such as steel grating and the control and counting room walls for lower energy $\gamma$-rays to be absorbed in and scatter from, processes more likely at lower photon energies. Therefore, the estimate of the rate in the OD is conservatively high, especially when considering that LZ construction in the cavern introduces more material for absorption and scattering.  
Combining this with the prediction for internal rate from LS contaminants in Ref. \cite{Haselschwardt:2018}, totalling $6 \pm 2$ Hz, and less than 12~Hz from simulation of radioactivity from LZ components (see Ref. \cite{LZ:2018}; an upper limit is given due to the use of upper limits in screening results for materials dominant to this rate), leads to a total rate in the LZ OD of $45\pm7$~Hz, significantly below the 100 Hz needed to maintain less than 5\% impact on LZ live time.
Furthermore, these results and the simulation model developed for this and other related studies has been used to inform the background model for a $^{136}$Xe $0\nu\beta\beta$-decay search in LZ, particularly influenced by the high energy $\gamma$-rays from the U and Th chains due to the Q-value of 2458~keV. %Current simulations predict that the cavern contributes more than 50\% of the background in a $2\sigma$ region of interest around the Q-value of 2458~keV.  

\section{Conclusion}
The $\gamma$-ray flux inside the Davis cavern at SURF has been measured using a sodium iodide detector, finding the corresponding radioactive contamination levels to be $220\pm60$~Bq/kg of $^{40}$K, $29\pm15$~Bq/kg of $^{238}$U, and $13\pm3$~Bq/kg of $^{232}$Th - consistent with shotcrete material used to coat the cavern walls. Radon in cavern air was found to produce a significant contribution to the measured rate, and there is no conclusive evidence for a significant or asymmetric flux from the high radioactivity rhyolite intrusion within the cavern. These results can be used to estimate the background contribution from the Davis cavern for the LZ dark matter experiment.  

\section*{Acknowledgements}
We would like to thank Keenan Thomas, now at Lawrence Livermore National Laboratory (LLNL), for his work on the HPGe measurements in the Davis Cavern. Additionally, we wish to thank the staff at SURF who helped with this measurement; in particular, those who were involved in tracking down a $3\frac{1}{2}$ inch floppy disk so that this data could actually leave the Davis Cavern! We are also grateful to the manufacturers of floppy disk-to-USB adaptors, making it possible (albeit through a further adaptor) to remove this data from the disk and analyse it. 

This work was partially supported by the U.S. Department of Energy (DOE) Office of Science under contract number DE-AC02-05CH11231 and under grant number DE-SC0019066; by the U.S. National Science Foundation (NSF); by the U.K. Science \& Technology Facilities Council under award numbers, ST/M003655/1, ST/M003981/1, ST/M003744/1, ST/M003639/1, ST/M003604/1, and ST/M003469/1; and by the Portuguese Foundation for Science and Technology (FCT) under award numbers CERN/FP/123610/2011 and PTDC/FISNUC/1525/2014; and by the Institute for Basic Science, Korea (budget numbers IBS-R016-D1, and IBS-R016-S1). University College London and Lawrence Berkeley National Laboratory thank the U.K. Royal Society for travel funds under the International Exchange Scheme (IE141517). We acknowledge additional support from the Boulby Underground Laboratory in the U.K.; the University of Wisconsin for grant UW PRJ82AJ; and the GridPP Collaboration, in particular at Imperial College London. This work was partially enabled by the University College London Cosmoparticle Initiative. Futhermore, this research used resources of the National Energy Research Scientific Computing Center, a DOE Office of Science User Facility supported by the Office of Science of the U.S. Department of Energy under Contract No. DE-AC02-05CH11231. We acknowledge many types of support provided to us by the South Dakota Science and Technology Authority (SDSTA), which developed the Sanford Underground Research Facility (SURF) with an important philanthropic donation from T. Denny Sanford as well as support from the State of South Dakota. SURF is operated by the SDSTA under contract to the Fermi National Accelerator Laboratory for the DOE, Office of Science. The University of Edinburgh is a charitable body, registered in Scotland, with the registration number SC005336. 
\newpage
\bibliographystyle{apsrev4-2}
\bibliography{main}

%apsrev4-2.bst 2015-08-30 from 4.21a (PWD, AO, DPC/HNN) hacked
\begin{thebibliography}{30}%
\makeatletter
\providecommand \@ifxundefined [1]{%
 \@ifx{#1\undefined}
}%
\providecommand \@ifnum [1]{%
 \ifnum #1\expandafter \@firstoftwo
 \else \expandafter \@secondoftwo
 \fi
}%
\providecommand \@ifx [1]{%
 \ifx #1\expandafter \@firstoftwo
 \else \expandafter \@secondoftwo
 \fi
}%
\providecommand \natexlab [1]{#1}%
\providecommand \enquote  [1]{``#1''}%
\providecommand \bibnamefont  [1]{#1}%
\providecommand \bibfnamefont [1]{#1}%
\providecommand \citenamefont [1]{#1}%
\providecommand \href@noop [0]{\@secondoftwo}%
\providecommand \href [0]{\begingroup \@sanitize@url \@href}%
\providecommand \@href[1]{\@@startlink{#1}\@@href}%
\providecommand \@@href[1]{\endgroup#1\@@endlink}%
\providecommand \@sanitize@url [0]{\catcode `\\12\catcode `\$12\catcode
  `\&12\catcode `\#12\catcode `\^12\catcode `\_12\catcode `\%12\relax}%
\providecommand \@@startlink[1]{}%
\providecommand \@@endlink[0]{}%
\providecommand \url  [0]{\begingroup\@sanitize@url \@url }%
\providecommand \@url [1]{\endgroup\@href {#1}{\urlprefix }}%
\providecommand \urlprefix  [0]{URL }%
\providecommand \Eprint [0]{\href }%
\providecommand \doibase [0]{http://dx.doi.org/}%
\providecommand \selectlanguage [0]{\@gobble}%
\providecommand \bibinfo  [0]{\@secondoftwo}%
\providecommand \bibfield  [0]{\@secondoftwo}%
\providecommand \translation [1]{[#1]}%
\providecommand \BibitemOpen [0]{}%
\providecommand \bibitemStop [0]{}%
\providecommand \bibitemNoStop [0]{.\EOS\space}%
\providecommand \EOS [0]{\spacefactor3000\relax}%
\providecommand \BibitemShut  [1]{\csname bibitem#1\endcsname}%
\let\auto@bib@innerbib\@empty
%</preamble>
\bibitem [{\citenamefont {Akerib}\ \emph {et~al.}(2017)\citenamefont {Akerib}
  \emph {et~al.}}]{Akerib:2016vxi}%
  \BibitemOpen
  \bibfield  {author} {\bibinfo {author} {\bibfnamefont {D.~S.}\ \bibnamefont
  {Akerib}} \emph {et~al.} (\bibinfo {collaboration}
  {\href{http://luxdarkmatter.org}{LUX}}),\ }\href {\doibase
  10.1103/PhysRevLett.118.021303} {\bibfield  {journal} {\bibinfo  {journal}
  {Phys. Rev. Lett.}\ }\textbf {\bibinfo {volume} {118}},\ \bibinfo {pages}
  {021303} (\bibinfo {year} {2017})},\ \Eprint
  {http://arxiv.org/abs/1608.07648}{arXiv:1608.07648 [astro-ph.CO]}\BibitemShut
  {NoStop}%
%%CITATION = ARXIV:1608.07648;%%
\bibitem [{\citenamefont {Cui}\ \emph {et~al.}(2017)\citenamefont {Cui} \emph
  {et~al.}}]{Cui:2017}%
  \BibitemOpen
  \bibfield  {author} {\bibinfo {author} {\bibfnamefont {X.}~\bibnamefont
  {Cui}} \emph {et~al.} (\bibinfo {collaboration}
  {\href{https://pandax.physics.sjtu.edu.cn}{PandaX}}),\ }\href {\doibase
  10.1103/PhysRevLett.119.181302} {\bibfield  {journal} {\bibinfo  {journal}
  {Phys. Rev. Lett.}\ }\textbf {\bibinfo {volume} {119}},\ \bibinfo {pages}
  {181302} (\bibinfo {year} {2017})}\BibitemShut {NoStop}%
\bibitem [{\citenamefont {Aprile}\ \emph {et~al.}(2018)\citenamefont {Aprile}
  \emph {et~al.}}]{Aprile:2018}%
  \BibitemOpen
  \bibfield  {author} {\bibinfo {author} {\bibfnamefont {E.}~\bibnamefont
  {Aprile}} \emph {et~al.} (\bibinfo {collaboration} {XENON1T}),\ }\href@noop
  {} {\enquote {\bibinfo {title} {{Dark Matter Search Results from a One
  Tonne$\times$Year Exposure of XENON1T}},}\ } (\bibinfo {year} {2018}),\
  \Eprint {http://arxiv.org/abs/1805.12562}{arXiv:1805.12562
  [astro-ph]}\BibitemShut {NoStop}%
%%CITATION = ARXIV:1805.12562;%%
\bibitem [{\citenamefont {Mount}\ \emph {et~al.}(2017)\citenamefont {Mount}
  \emph {et~al.}}]{mount:2017}%
  \BibitemOpen
  \bibfield  {author} {\bibinfo {author} {\bibfnamefont {B.~J.}\ \bibnamefont
  {Mount}} \emph {et~al.} (\bibinfo {collaboration}
  {\href{http://lz.lbl.gov}{LZ}}),\ }\href@noop {} {\bibfield  {journal}
  {\bibinfo  {journal} {{}}\ } (\bibinfo {year} {2017})},\ \bibinfo {note}
  {{Technical Design Report; LBNL-1007256, FERMILAB-TM-2653-AE-E-PPD}},\
  \Eprint {http://arxiv.org/abs/1703.09144}{arXiv:1703.09144
  [physics.ins-det]}\BibitemShut {NoStop}%
%%CITATION = ARXIV:1703.09144;%%
\bibitem [{\citenamefont {Akerib}\ \emph {et~al.}(2018)\citenamefont {Akerib}
  \emph {et~al.}}]{LZ:2018}%
  \BibitemOpen
  \bibfield  {author} {\bibinfo {author} {\bibfnamefont {D.}~\bibnamefont
  {Akerib}} \emph {et~al.} (\bibinfo {collaboration}
  {\href{http://lz.lbl.gov}{LZ}}),\ }\href@noop {} {\  (\bibinfo {year}
  {2018})},\ \Eprint {http://arxiv.org/abs/1802.06039}{arXiv:1802.06039
  [astro-ph]}\BibitemShut {NoStop}%
%%CITATION = ARXIV:1802.06039 ;%%
\bibitem [{\citenamefont {Akerib}\ \emph
  {et~al.}(2016{\natexlab{a}})\citenamefont {Akerib} \emph
  {et~al.}}]{akerib:2017}%
  \BibitemOpen
  \bibfield  {author} {\bibinfo {author} {\bibfnamefont {D.}~\bibnamefont
  {Akerib}} \emph {et~al.} (\bibinfo {collaboration} {LZ}),\ }\href@noop {}
  {\bibfield  {journal} {\bibinfo  {journal} {Astropart. Phys.}\ }\textbf
  {\bibinfo {volume} {96}},\ \bibinfo {pages} {1} (\bibinfo {year}
  {2016}{\natexlab{a}})}\BibitemShut {NoStop}%
\bibitem [{\citenamefont {Lesko}(2015)}]{lesko:2015sma}%
  \BibitemOpen
  \bibfield  {author} {\bibinfo {author} {\bibfnamefont {K.~T.}\ \bibnamefont
  {Lesko}},\ }\bibfield  {booktitle} {\emph {\bibinfo {booktitle}
  {{Proceedings, 13th International Conference on Topics in Astroparticle and
  Underground Physics (TAUP 2013): Asilomar, California, September 8-13,
  2013}}},\ }\href {\doibase 10.1016/j.phpro.2014.12.001} {\bibfield  {journal}
  {\bibinfo  {journal} {Phys. Procedia}\ }\textbf {\bibinfo {volume} {61}},\
  \bibinfo {pages} {542} (\bibinfo {year} {2015})}\BibitemShut {NoStop}%
%%CITATION = INSPIRE-1357738;%%
\bibitem [{\citenamefont {Akerib}\ \emph
  {et~al.}(2016{\natexlab{b}})\citenamefont {Akerib} \emph
  {et~al.}}]{Akerib:2016lao}%
  \BibitemOpen
  \bibfield  {author} {\bibinfo {author} {\bibfnamefont {D.~S.}\ \bibnamefont
  {Akerib}} \emph {et~al.} (\bibinfo {collaboration}
  {\href{http://luxdarkmatter.org}{LUX}}),\ }\href {\doibase
  10.1103/PhysRevLett.116.161302} {\bibfield  {journal} {\bibinfo  {journal}
  {Phys. Rev. Lett.}\ }\textbf {\bibinfo {volume} {116}},\ \bibinfo {pages}
  {161302} (\bibinfo {year} {2016}{\natexlab{b}})},\ \Eprint
  {http://arxiv.org/abs/1602.03489}{arXiv:1602.03489 [hep-ex]}\BibitemShut
  {NoStop}%
%%CITATION = ARXIV:1602.03489;%%
\bibitem [{\citenamefont {Cherry}\ \emph {et~al.}(1983)\citenamefont {Cherry},
  \citenamefont {Deakyne}, \citenamefont {Lande}, \citenamefont {Lee},
  \citenamefont {Steinberg}, \citenamefont {Cleveland},\ and\ \citenamefont
  {Fenyves}}]{cherry:1983dp}%
  \BibitemOpen
  \bibfield  {author} {\bibinfo {author} {\bibfnamefont {M.~L.}\ \bibnamefont
  {Cherry}}, \bibinfo {author} {\bibfnamefont {M.}~\bibnamefont {Deakyne}},
  \bibinfo {author} {\bibfnamefont {K.}~\bibnamefont {Lande}}, \bibinfo
  {author} {\bibfnamefont {C.~K.}\ \bibnamefont {Lee}}, \bibinfo {author}
  {\bibfnamefont {R.~I.}\ \bibnamefont {Steinberg}}, \bibinfo {author}
  {\bibfnamefont {B.~T.}\ \bibnamefont {Cleveland}}, \ and\ \bibinfo {author}
  {\bibfnamefont {E.~J.}\ \bibnamefont {Fenyves}},\ }\href {\doibase
  10.1103/PhysRevD.27.1444} {\bibfield  {journal} {\bibinfo  {journal} {Phys.
  Rev.}\ }\textbf {\bibinfo {volume} {D27}},\ \bibinfo {pages} {1444} (\bibinfo
  {year} {1983})}\BibitemShut {NoStop}%
%%CITATION = PHRVA,D27,1444;%%
\bibitem [{\citenamefont {Gray}\ \emph {et~al.}(2011)\citenamefont {Gray},
  \citenamefont {Ruybal}, \citenamefont {Totushek}, \citenamefont {Mei},
  \citenamefont {Thomas},\ and\ \citenamefont {Zhang}}]{gray:2010nc}%
  \BibitemOpen
  \bibfield  {author} {\bibinfo {author} {\bibfnamefont {F.~E.}\ \bibnamefont
  {Gray}}, \bibinfo {author} {\bibfnamefont {C.}~\bibnamefont {Ruybal}},
  \bibinfo {author} {\bibfnamefont {J.}~\bibnamefont {Totushek}}, \bibinfo
  {author} {\bibfnamefont {D.-M.}\ \bibnamefont {Mei}}, \bibinfo {author}
  {\bibfnamefont {K.}~\bibnamefont {Thomas}}, \ and\ \bibinfo {author}
  {\bibfnamefont {C.}~\bibnamefont {Zhang}},\ }\href {\doibase
  10.1016/j.nima.2011.02.032} {\bibfield  {journal} {\bibinfo  {journal} {Nucl.
  Instrum. Meth.}\ }\textbf {\bibinfo {volume} {A638}},\ \bibinfo {pages} {63}
  (\bibinfo {year} {2011})},\ \Eprint
  {http://arxiv.org/abs/1007.1921}{arXiv:1007.1921 [nucl-ex]}\BibitemShut
  {NoStop}%
%%CITATION=ARXIV:1007.1921;%%
\bibitem [{\citenamefont {Abgrall}\ \emph {et~al.}(2017)\citenamefont {Abgrall}
  \emph {et~al.}}]{majorana:2017}%
  \BibitemOpen
  \bibfield  {author} {\bibinfo {author} {\bibfnamefont {N.}~\bibnamefont
  {Abgrall}} \emph {et~al.} (\bibinfo {collaboration}
  {\href{https://www.npl.washington.edu/majorana/majorana-experiment}{Majorana}}),\
  }\href@noop {} {\bibfield  {journal} {\bibinfo  {journal} {Astropart. Phys}\
  }\textbf {\bibinfo {volume} {93}} (\bibinfo {year} {2017})},\ \Eprint
  {http://arxiv.org/abs/1602.07742}{1602.07742}\BibitemShut {NoStop}%
\bibitem [{\citenamefont {Akerib}\ \emph {et~al.}(2013)\citenamefont {Akerib}
  \emph {et~al.}}]{akerib:2012ys}%
  \BibitemOpen
  \bibfield  {author} {\bibinfo {author} {\bibfnamefont {D.~S.}\ \bibnamefont
  {Akerib}} \emph {et~al.} (\bibinfo {collaboration}
  {\href{http://luxdarkmatter.org}{LUX}}),\ }\href {\doibase
  10.1016/j.nima.2012.11.135} {\bibfield  {journal} {\bibinfo  {journal} {Nucl.
  Instrum. Meth.}\ }\textbf {\bibinfo {volume} {A704}},\ \bibinfo {pages} {111}
  (\bibinfo {year} {2013})},\ \Eprint
  {http://arxiv.org/abs/1211.3788}{arXiv:1211.3788
  [physics.ins-det]}\BibitemShut {NoStop}%
%%CITATION = ARXIV:1211.3788;%%
\bibitem [{\citenamefont {Hart}(2010)}]{hart:2010}%
  \BibitemOpen
  \bibfield  {author} {\bibinfo {author} {\bibfnamefont {K.}~\bibnamefont
  {Hart}},\ }\href@noop {} {}\bibinfo {howpublished} {private communication}
  (\bibinfo {year} {2010})\BibitemShut {NoStop}%
\bibitem [{\citenamefont {Heise}(2014)}]{heise:2015}%
  \BibitemOpen
  \bibfield  {author} {\bibinfo {author} {\bibfnamefont {J.}~\bibnamefont
  {Heise}},\ }in\ \href@noop {} {\emph {\bibinfo {booktitle} {Proceedings of
  the Workshop on Germanium-Based Detectors and Technology}}}\ (\bibinfo {year}
  {2014})\ \bibinfo {note} {\href{https://arxiv.org/abs/1503.01112}{The Sanford
  Underground Research Facility at Homestake}}\BibitemShut {NoStop}%
\bibitem [{\citenamefont {Kinsey}\ \emph {et~al.}(1996)\citenamefont {Kinsey}
  \emph {et~al.}}]{nudat:1996}%
  \BibitemOpen
  \bibfield  {author} {\bibinfo {author} {\bibfnamefont {R.~R.}\ \bibnamefont
  {Kinsey}} \emph {et~al.},\ }\href@noop {} {\emph {\bibinfo {title} {9th
  International Symposium of Capture Gamma-Ray Spectroscopy and Related Topics,
  Budapest, Hungary}}}\ (\bibinfo {year} {1996})\ \bibinfo {note}
  {\href{http://www.nndc.bnl.gov/nudat2/}{ Data extracted from the NUDAT
  database, version 2.7, 2018}}\BibitemShut {NoStop}%
\bibitem [{\citenamefont {Malczewski}\ \emph
  {et~al.}(2012{\natexlab{a}})\citenamefont {Malczewski}, \citenamefont
  {Kisiel},\ and\ \citenamefont {Dorda}}]{gransasso:2012}%
  \BibitemOpen
  \bibfield  {author} {\bibinfo {author} {\bibfnamefont {D.}~\bibnamefont
  {Malczewski}}, \bibinfo {author} {\bibfnamefont {J.}~\bibnamefont {Kisiel}},
  \ and\ \bibinfo {author} {\bibfnamefont {J.}~\bibnamefont {Dorda}},\ }\href
  {\doibase 10.1007/s10967-012-1990-9} {\bibfield  {journal} {\bibinfo
  {journal} {J. Radioanal Nucl Chem}\ }\textbf {\bibinfo {volume} {295}},\
  \bibinfo {pages} {751} (\bibinfo {year} {2012}{\natexlab{a}})}\BibitemShut
  {NoStop}%
\bibitem [{\citenamefont {Malczewski}\ \emph
  {et~al.}(2012{\natexlab{b}})\citenamefont {Malczewski}, \citenamefont
  {Kisiel},\ and\ \citenamefont {Dorda}}]{modane:2012}%
  \BibitemOpen
  \bibfield  {author} {\bibinfo {author} {\bibfnamefont {D.}~\bibnamefont
  {Malczewski}}, \bibinfo {author} {\bibfnamefont {J.}~\bibnamefont {Kisiel}},
  \ and\ \bibinfo {author} {\bibfnamefont {J.}~\bibnamefont {Dorda}},\ }\href
  {\doibase 10.1007/s10967-011-1497-9} {\bibfield  {journal} {\bibinfo
  {journal} {J. Radioanal Nucl Chem}\ }\textbf {\bibinfo {volume} {292}},\
  \bibinfo {pages} {751} (\bibinfo {year} {2012}{\natexlab{b}})}\BibitemShut
  {NoStop}%
\bibitem [{\citenamefont {Malczewski}\ \emph {et~al.}(2013)\citenamefont
  {Malczewski}, \citenamefont {Kisiel},\ and\ \citenamefont
  {Dorda}}]{boulby:2013}%
  \BibitemOpen
  \bibfield  {author} {\bibinfo {author} {\bibfnamefont {D.}~\bibnamefont
  {Malczewski}}, \bibinfo {author} {\bibfnamefont {J.}~\bibnamefont {Kisiel}},
  \ and\ \bibinfo {author} {\bibfnamefont {J.}~\bibnamefont {Dorda}},\ }\href
  {\doibase 10.1007/s10967-013-254909} {\bibfield  {journal} {\bibinfo
  {journal} {J. Radioanal Nucl Chem}\ }\textbf {\bibinfo {volume} {298}},\
  \bibinfo {pages} {1483} (\bibinfo {year} {2013})}\BibitemShut {NoStop}%
\bibitem [{\citenamefont {Zeng}\ \emph {et~al.}(2014)\citenamefont {Zeng} \emph
  {et~al.}}]{jinping:2014}%
  \BibitemOpen
  \bibfield  {author} {\bibinfo {author} {\bibfnamefont {Z.}~\bibnamefont
  {Zeng}} \emph {et~al.},\ }\href {\doibase 10.1007/s10967-014-3114-1}
  {\bibfield  {journal} {\bibinfo  {journal} {J. Radioanal Nucl Chem}\ }\textbf
  {\bibinfo {volume} {301}},\ \bibinfo {pages} {443} (\bibinfo {year}
  {2014})}\BibitemShut {NoStop}%
\bibitem [{\citenamefont {Smith}(2007)}]{smith:2007}%
  \BibitemOpen
  \bibfield  {author} {\bibinfo {author} {\bibfnamefont {A.}~\bibnamefont
  {Smith}},\ }\href@noop {} {\emph {\bibinfo {title} {{Homestake (DUSEL)
  samples — results of radiometric analyses at LBNL}}}},\ \bibinfo {type}
  {Tech. Rep.}\ (\bibinfo  {institution} {Lawrence Berkeley National Laboratory
  (LBNL), 1 Cyclotron Road, Berkeley, CA 94720-8099, USA},\ \bibinfo {year}
  {2007})\BibitemShut {NoStop}%
\bibitem [{\citenamefont {Mei}\ \emph {et~al.}(2010)\citenamefont {Mei},
  \citenamefont {Zhang}, \citenamefont {Thomas},\ and\ \citenamefont
  {Gray}}]{mei:2009py}%
  \BibitemOpen
  \bibfield  {author} {\bibinfo {author} {\bibfnamefont {D.-M.}\ \bibnamefont
  {Mei}}, \bibinfo {author} {\bibfnamefont {C.}~\bibnamefont {Zhang}}, \bibinfo
  {author} {\bibfnamefont {K.}~\bibnamefont {Thomas}}, \ and\ \bibinfo {author}
  {\bibfnamefont {F.}~\bibnamefont {Gray}},\ }\href {\doibase
  10.1016/j.astropartphys.2010.04.003} {\bibfield  {journal} {\bibinfo
  {journal} {Astropart. Phys.}\ }\textbf {\bibinfo {volume} {34}},\ \bibinfo
  {pages} {33} (\bibinfo {year} {2010})},\ \Eprint
  {http://arxiv.org/abs/0912.0211}{arXiv:0912.0211 [nucl-ex]}\BibitemShut
  {NoStop}%
%%CITATION=ARXIV:0912.0211;%%
\bibitem [{\citenamefont {{ORTEC}}()}]{MAESTRO}%
  \BibitemOpen
  \bibfield  {author} {\bibinfo {author} {\bibnamefont {{ORTEC}}},\ }\href
  {https://www.ortec-online.com/products/application-software/maestro-mca}
  {\enquote {\bibinfo {title} {{MAESTRO Multichannel Analyzer Emulation}},}\
  }\bibinfo {note}
  {{https://www.ortec-online.com/products/application-software/maestro-mca}}\BibitemShut
  {NoStop}%
\bibitem [{\citenamefont {An}\ \emph {et~al.}(2017)\citenamefont {An} \emph
  {et~al.}}]{an:2017}%
  \BibitemOpen
  \bibfield  {author} {\bibinfo {author} {\bibfnamefont {F.~P.}\ \bibnamefont
  {An}} \emph {et~al.} (\bibinfo {collaboration}
  {\href{http://dayabay.ihep.ac.cn/twiki/bin/view/Public/}{Daya Bay}}),\ }\href
  {\doibase 10.1103/PhysRevD.95.072006} {\bibfield  {journal} {\bibinfo
  {journal} {Phys. Rev. D.}\ }\textbf {\bibinfo {volume} {95}},\ \bibinfo
  {pages} {072006} (\bibinfo {year} {2017})}\BibitemShut {NoStop}%
\bibitem [{\citenamefont {Adhikari}\ \emph {et~al.}(2016)\citenamefont
  {Adhikari} \emph {et~al.}}]{Adhikari:2016}%
  \BibitemOpen
  \bibfield  {author} {\bibinfo {author} {\bibnamefont {Adhikari}} \emph
  {et~al.},\ }\href@noop {} {\bibfield  {journal} {\bibinfo  {journal} {Eur.
  Phys. J.}\ }\textbf {\bibinfo {volume} {C76}},\ \bibinfo {pages} {185}
  (\bibinfo {year} {2016})}\BibitemShut {NoStop}%
\bibitem [{\citenamefont {Agostinelli}\ \emph {et~al.}(2003)\citenamefont
  {Agostinelli} \emph {et~al.}}]{Agostinelli:2002hh}%
  \BibitemOpen
  \bibfield  {author} {\bibinfo {author} {\bibfnamefont {S.}~\bibnamefont
  {Agostinelli}} \emph {et~al.} (\bibinfo {collaboration}
  {\href{http://www.geant4.org/geant4/}{GEANT4}}),\ }\href {\doibase
  10.1016/S0168-9002(03)01368-8} {\bibfield  {journal} {\bibinfo  {journal}
  {Nucl. Instrum. Meth.}\ }\textbf {\bibinfo {volume} {A506}},\ \bibinfo
  {pages} {250} (\bibinfo {year} {2003})}\BibitemShut {NoStop}%
%%CITATION=NUIMA,A506,250;%%
\bibitem [{\citenamefont {Cullen}\ \emph {et~al.}(1997)\citenamefont {Cullen},
  \citenamefont {Hubbell},\ and\ \citenamefont {Kissel}}]{liv1}%
  \BibitemOpen
  \bibfield  {author} {\bibinfo {author} {\bibfnamefont {D.~E.}\ \bibnamefont
  {Cullen}}, \bibinfo {author} {\bibfnamefont {J.~H.}\ \bibnamefont {Hubbell}},
  \ and\ \bibinfo {author} {\bibfnamefont {L.}~\bibnamefont {Kissel}},\ }\href
  {\doibase 10.2172/295438} {\emph {\bibinfo {title} {EPDL97: the evaluated
  photo data library `97 version}}},\ \bibinfo {type} {Tech. Rep.}\ (\bibinfo
  {year} {1997})\BibitemShut {NoStop}%
\bibitem [{\citenamefont {Perkins}\ \emph {et~al.}(1991)\citenamefont
  {Perkins}, \citenamefont {Cullen}, \citenamefont {Seltzer},\ and\
  \citenamefont {Technology~(NML)}}]{liv2}%
  \BibitemOpen
  \bibfield  {author} {\bibinfo {author} {\bibfnamefont {S.~T.}\ \bibnamefont
  {Perkins}}, \bibinfo {author} {\bibfnamefont {C.~U.~S.}\ \bibnamefont
  {Cullen}, \bibfnamefont {D.~E. (Lawrence Livermore National~Lab.}}, \bibinfo
  {author} {\bibfnamefont {S.~M. N. I. o.~S.}\ \bibnamefont {Seltzer}}, \ and\
  \bibinfo {author} {\bibfnamefont {M.~U. S. C. f. R.~R.}\ \bibnamefont
  {Technology~(NML)}, \bibfnamefont {Gaithersburg}},\ }\href {\doibase
  10.2172/5691165} {\emph {\bibinfo {title} {Tables and graphs of
  electron-interaction cross sections from 10 eV to 100 GeV derived from the
  LLNL Evaluated Electron Data Library (EEDL), Z = 1--100}}},\ \bibinfo {type}
  {Tech. Rep.}\ (\bibinfo {year} {1991})\BibitemShut {NoStop}%
\bibitem [{\citenamefont {Tomasello}\ \emph {et~al.}(2010)\citenamefont
  {Tomasello}, \citenamefont {Robinson},\ and\ \citenamefont
  {Kudryavtsev}}]{tomasello:2010zz}%
  \BibitemOpen
  \bibfield  {author} {\bibinfo {author} {\bibfnamefont {V.}~\bibnamefont
  {Tomasello}}, \bibinfo {author} {\bibfnamefont {M.}~\bibnamefont {Robinson}},
  \ and\ \bibinfo {author} {\bibfnamefont {V.~A.}\ \bibnamefont
  {Kudryavtsev}},\ }\href {\doibase 10.1016/j.astropartphys.2010.05.005}
  {\bibfield  {journal} {\bibinfo  {journal} {Astropart. Phys.}\ }\textbf
  {\bibinfo {volume} {34}},\ \bibinfo {pages} {70} (\bibinfo {year}
  {2010})}\BibitemShut {NoStop}%
%%CITATION = APHYE,34,70;%%
\bibitem [{\citenamefont {Thomas}(2014)}]{thomas:2014}%
  \BibitemOpen
  \bibfield  {author} {\bibinfo {author} {\bibfnamefont {K.~J.}\ \bibnamefont
  {Thomas}} (\bibinfo {collaboration} {\href{http://lz.lbl.gov}{LZ}}),\
  }\href@noop {} {\emph {\bibinfo {title}
  {\href{http://hep.ucsb.edu/LZ/BLBF-Davis-gammaflux.pdf}{An Estimate of the
  Gamma Flux in the East Counting Room of the Davis Cavern}}}},\ \bibinfo
  {type} {Tech. Rep.}\ (\bibinfo  {institution} {Lawrence Berkeley National
  Laboratory (LBNL), 1 Cyclotron Road, Berkeley, CA 94720-8099, USA},\ \bibinfo
  {year} {2014})\BibitemShut {NoStop}%
\bibitem [{\citenamefont {Haselschwardt}\ \emph {et~al.}(2019)\citenamefont
  {Haselschwardt}, \citenamefont {Shaw}, \citenamefont {Nelson}, \citenamefont
  {Witherell}, \citenamefont {Yeh}, \citenamefont {Lesko}, \citenamefont
  {Cole}, \citenamefont {Kyre},\ and\ \citenamefont
  {White}}]{Haselschwardt:2018}%
  \BibitemOpen
  \bibfield  {author} {\bibinfo {author} {\bibfnamefont {S.}~\bibnamefont
  {Haselschwardt}}, \bibinfo {author} {\bibfnamefont {S.}~\bibnamefont {Shaw}},
  \bibinfo {author} {\bibfnamefont {H.}~\bibnamefont {Nelson}}, \bibinfo
  {author} {\bibfnamefont {M.}~\bibnamefont {Witherell}}, \bibinfo {author}
  {\bibfnamefont {M.}~\bibnamefont {Yeh}}, \bibinfo {author} {\bibfnamefont
  {K.}~\bibnamefont {Lesko}}, \bibinfo {author} {\bibfnamefont
  {A.}~\bibnamefont {Cole}}, \bibinfo {author} {\bibfnamefont {S.}~\bibnamefont
  {Kyre}}, \ and\ \bibinfo {author} {\bibfnamefont {D.}~\bibnamefont {White}},\
  }\href@noop {} {\bibfield  {journal} {\bibinfo  {journal} {Nucl. Instrum.
  Meth.}\ }\textbf {\bibinfo {volume} {A937}},\ \bibinfo {pages} {148}
  (\bibinfo {year} {2019})}\BibitemShut {NoStop}%
\end{thebibliography}%

\end{document}